\documentclass[10pt,aps,pra,twocolumn,superscriptaddress,floatfix,nofootinbib]{revtex4-2}

\usepackage[english]{babel}
\usepackage{soul}
\usepackage[T1]{fontenc}
\usepackage{amsfonts,amsmath}
\newtheorem{definition}{Definition}
\usepackage{siunitx}
\usepackage{braket}
\usepackage{mathrsfs}
\usepackage{mathtools}
\usepackage{booktabs}
\usepackage{multirow}
\usepackage{lipsum} 
\usepackage{varwidth}
\usepackage{standalone}
\usepackage{moresize}
\usepackage{placeins}
\usepackage{float}

\usepackage[breaklinks=true]{hyperref}
\usepackage[usenames,dvipsnames]{color}
\hypersetup{
  colorlinks   = true, 
  urlcolor     = blue, 
  linkcolor    = blue, 
  citecolor   = red 
}
\usepackage[noabbrev]{cleveref}

\usepackage{tikz}
\usetikzlibrary{quantikz}
\usepackage{circuitikz}

\usetikzlibrary{positioning}
\usetikzlibrary{calc}

\tikzset{
  pics/jjbox/.style={
    code={
      \draw[line width=1.1pt] (-0.65,0) -- (-0.25,0);
      \draw[line width=1.1pt] ( 0.25,0) -- ( 0.65,0);

      \draw[line width=1.1pt]
        (-0.25,-0.25) rectangle (0.25,0.25);

      \draw[line width=1.1pt]
        (-0.25,-0.25) -- (0.25,0.25);
      \draw[line width=1.1pt]
        (-0.25, 0.25) -- (0.25,-0.25);
    }
  },
  pics/jj/.style={
    code={
      \draw[line width=1.1pt] (-0.65,0) -- (0,0);
      \draw[line width=1.1pt] ( 0.0,0) -- ( 0.65,0);

      \draw[line width=1.1pt]
        (-0.25,-0.25) -- (0.25,0.25);
      \draw[line width=1.1pt]
        (-0.25, 0.25) -- (0.25,-0.25);
    }
  }
}

\usepackage{algorithm}
\usepackage{algpseudocode}

\newcommand{\zp}{$0$-$\pi$}

\crefformat{equation}{Eq.~(#2#1#3)} 
\crefformat{section}{Sec.~#2#1#3} 
\Crefformat{equation}{Equation~(#2#1#3)}
\crefformat{figure}{Fig.~#2#1#3}
\crefrangeformat{equation}{Eqs.~#3(#1)#4--#5(#2)#6}
\crefmultiformat{equation}{Eqs.~(#2#1#3)}{ and~(#2#1#3)}{, (#2#1#3)}{ and~(#2#1#3)}

\newcommand{\canonical}{reference}
\newcommand{\Canonical}{Reference}

\makeatother

\begin{document}

\hfuzz=150pt
\hbadness=10000

\title{
Enumeration of all superconducting circuits up to 5 nodes
}
\author{Eli J. Weissler}
\email[]{eli.weissler@colorado.edu}
\affiliation{Electrical, Computer, and Energy Engineering, University of Colorado Boulder, Colorado 80309, USA}
\author{Mohit Bhat}
\affiliation{Electrical, Computer, and Energy Engineering, University of Colorado Boulder, Colorado 80309, USA}
\author{Zhenxing Liu}
\affiliation{Beijing Key Laboratory of Fault-Tolerant Quantum Computing, Beijing Academy of Quantum Information Sciences, Beijing 100193, China}
\affiliation{Materials Science and Engineering, University of Colorado Boulder, Colorado 80309, USA}
\author{Joshua Combes}
\affiliation{Electrical, Computer, and Energy Engineering, University of Colorado Boulder, Colorado 80309, USA}
\affiliation{School of Physics and School of Mathematics, The University of Melbourne, VIC 3010, Australia}

\date{September 17, 2026}

\begin{abstract}
Nonlinear superconducting circuits can be used as amplifiers, transducers, and qubits. Only a handful have been analyzed or built, so useful configurations likely remain undiscovered. In this paper, we enumerate all unique nonlinear superconducting circuits -- up to five nodes in size -- built of capacitors, inductors, and Josephson junctions. 
We catalog the resulting design space by sorting circuits into ``Hamiltonian classes'' using a set of \canonical~coordinates.
Finally, we search for novel superconducting qubits by explicitly considering all three-node circuits, showing how the results of our enumeration can be used as a starting point for circuit design tasks.
\end{abstract}
\maketitle

\section{Introduction}

Superconducting quantum circuits have gained widespread use over the last 25 years as qubits, amplifiers, and transducers~\cite{BlasGrimGirvWallr2021}. However, most devices have been built from a small library of circuits. The most widely used superconducting qubit today, the transmon, consists of a Josephson junction in parallel with a capacitor. Large-scale industrial efforts in superconducting quantum computing rely on this basic design, building complex systems by repeating the transmon circuit. With the advancements of the last two decades, we are now in a position to design larger multimode circuits, which are necessary to realize systems with intrinsic protection against both decay and dephasing \cite{Gyenis2021}.

\begin{figure}[t]
\centering 
\includegraphics[width=0.99\columnwidth]{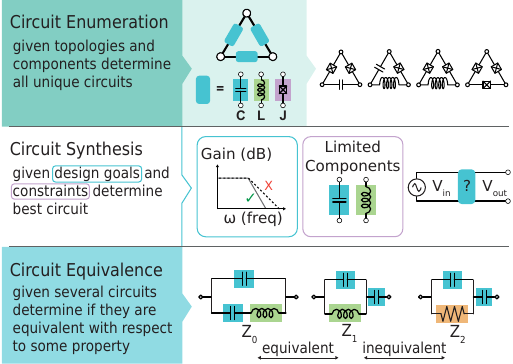}
\caption{Visual representations of circuit enumeration, synthesis, and equivalence. The aim of our work is to enumerate all unique nonlinear quantum circuits with up to five nodes. The results of our enumeration leverage circuit equivalence and can be used in circuit synthesis tasks to find new qubits and amplifiers. In the circuit synthesis example, we desire a low-pass filter with a certain roll-off and search for an LC circuit that has this property. The illustrated circuit equivalence is defined with respect to the impedance $Z_i$.
}
\label{fig:enumeration_vs_synthesis_vs_equiv}
\end{figure}

 Classical nonlinear circuits have been studied for over a century~\cite{fleming_1905,forest_1906}. Large, complex analog and digital circuits are constructed by connecting smaller well-understood modules~\cite{sedra2020}. 
 These smaller circuits or modules, usually containing fewer than 10 nonlinear elements, are designed for specific functions. 
In the discovery and design of these modules, three key concepts are closely connected (depicted in \cref{fig:enumeration_vs_synthesis_vs_equiv}):
\begin{enumerate}
    \item \textsc{circuit enumeration} uses a combinatorial approach to list all unique circuit configurations, given a set of components. 
    \item \textsc{circuit synthesis} constructs a circuit that meets performance requirements, subject to constraints.  
    \item \textsc{circuit equivalence} determines if different circuits perform the identical function.
\end{enumerate}
In circuit design, the goal is to synthesize high-performing circuits. This process becomes inefficient if an algorithm suggests circuits equivalent to previously examined ones or cannot explore new topologies.

This paper enumerates all superconducting quantum circuits with up to five nodes built from capacitors, inductors, and Josephson junctions.\footnote{It is straightforward to include other two-terminal devices such as the quantum phase-slip junction, but it is difficult to include three- or more-terminal devices; see \cref{subsec:elems considered} for 
more details.}  We consider circuits up to five nodes in size due to computational constraints, though the procedure could be applied to limited subsets of larger circuits. Our enumeration framework has two related components. First, circuit enumeration maps the design space and identifies equivalent circuits, providing a systematic basis for circuit synthesis. Second, we group the quantized circuits into Hamiltonian classes according to their mode and coupling structure. As a proof of concept for enumeration, we search all unique three-node circuits for new superconducting qubits. As a proof of concept for the Hamiltonian classification, we start from a target Hamiltonian and then identify all circuits that realize it, and we use the classification to find circuits whose nonlinear coordinates are necessarily delocalized.

The remainder of this paper is structured as follows. In \cref{sec:History}, we give brief historical context for circuit enumeration and summarize related work. In \cref{sec:CirEnumAlg}, we describe our circuit enumeration algorithm, which leans on the work of Herber et al.~\cite{Herber:2017aa, herber:2017ab, herber_enhancements_2020} to create a set of topologically unique circuits. In \cref{sec:HamEnum}, we compute the Hamiltonian of each unique circuit and categorize the circuits into Hamiltonian equivalence classes relative to a particular choice of \canonical\ coordinates.
In \cref{sec:results}, we present the enumeration and classification results and emphasize that distinct circuits can belong to the same Hamiltonian class.
In \cref{sec:optimization}, we develop a numerical procedure to estimate and optimize circuit performance as a qubit, then apply it to all unique three-node circuits.
This numerical work leverages several existing circuit quantization software packages~\cite{Groszkowski2021_scqubitspython,chitta2022_scqubits,kqcircuits,Gely2020_QuCAT,Aumann2022_CircuitQ,Rajabzadeh2023analysisofarbitrary,rajabzadeh_2024, Tanamoto_2023}. Finally, in \cref{sec:conc}, we conclude with a discussion of open problems.

\section{Historical Perspective \& Related work}\label{sec:History}

In the late 1800s, circuit enumeration became important due to the rapid expansion of telephone networks.
As Darlington recounts~\cite{Darlington:1984aa}:
Foster and Campbell~\cite{Campbell:1920aa} were exploring four-port transformer circuits and initially aimed to patent several promising designs. However, Bell Labs advised against patenting designs for which equivalent circuits existed. Upon realizing the vast number of possible circuits (83,539), they instead published the enumeration, which kept these designs open for use.

Circuit enumeration has a long and active history~\cite{Kirchhoff1847a,Kirchhoff1847b,MacMahon:1890aa,Howitt:1931aa,Brune:1931aa,Brune:1931ab,Foster:1932aa,Tellegen:1940aa,RiordanShannon:1942aa,Carlitz:1956aa,Belevitch:1962aa,Mateti:1976aa,Watanabe:1999aa,Herber:2017aa,herber:2017ab,Lohan2019:aa,Morelli2019:aa}. Most work has focused on circuits with two-terminal, passive linear elements, 
although there has been progress on nonlinear and multiport devices~\cite{Happ1967,Perelson76aa,Reibiger:2013aa}. For example, one set of studies identified previously unknown amplifier circuits~\cite{Bruccoleri2001,Klumperink2001,Shahhosseini2018,Pretl2021}. Other work has explored the integration of machine learning and advanced optimization methods into more traditional approaches to enumeration and synthesis~\cite{Lyu2018,guo_circuit_2019,Zhao2022,Fayazi2023,sirico_use_2023}, with a recent focus on non-redundant representations of the circuit design space~\cite{Gao2025AnalogGenie}.

The systematic study of quantum circuit theory, now often called circuit quantum electrodynamics (circuit QED)~\cite{BlasGrimGirvWallr2021}, was initiated by \citet{YurkeDenker1984}. 
This body of work is largely disconnected from classical circuit theory, as analyzing quantum circuits requires the use of Lagrangian and Hamiltonian mechanics, a formalism rarely used in classical electrical engineering~\cite{Chua1974,Bernstein1989,Weiss1997}.
The field of circuit QED has grown rapidly, driven by experimental breakthroughs in the early 2000s~\cite{BouchiatDevoret1998,NakamuraPashkin1999,VionDevoret2002,MartinisNam2002,ChiorescuMooij2003,WallraffSchoelkopf2004} and by continued progress over the past two decades~\cite{Schuster2007,krinner_realizing_2022,google_surface_2024}.

The field has mainly focused on superconducting qubits, driven by their importance in quantum computing. Over the past few decades, the most widely used superconducting-qubit circuit has been a Josephson junction in parallel with a capacitor, namely the Cooper-pair box or transmon~\cite{BouchiatDevoret1998,Koch2007}. The transition from the Cooper-pair box to the transmon kept the circuit unchanged but moved to a parameter regime with much lower noise.
By optimizing the capacitance and Josephson energy, coherence times have increased from nanoseconds~\cite{BouchiatDevoret1998}  to median values of $T_1$ and $T_2^{\rm echo}$ of about $0.5$ milliseconds~\cite{tuokkola_methods_2024}. Of course, improvements in materials and fabrication have also contributed substantially to this increase~\cite{place_new_2021, wang_towards_2022,tuokkola_methods_2024}.

The majority of superconducting quantum circuits built today are small two-node circuits made of capacitors (\verb|C|), inductors (\verb|L|), and Josephson junctions (\verb|J|). Important examples include the resonator (\verb|LC|), the transmon (\verb|JC|), and fluxonium (\verb|JCL|)~\cite{Manucharyan2009,SomoroffManucharyan2023}. While these circuits are often modeled as lumped elements, their physical realization is more complex. For example, the inductor in fluxonium is typically constructed from an array of Josephson junctions, which behaves like a linear inductor only at certain frequencies \cite{manucharyan_thesis:2012}.\footnote{Developing large inductors, not built from junction arrays, is an active area of research~\cite{grunhaupt_granular_2019, peruzzo_surpassing_2020, gupta_low_2024}.} In recent years, larger multimode circuits have gained interest, driven primarily by the subfield of ``protected qubits''~\cite{Gyenis2021}. Identifying the most promising multimode superconducting systems remains an open question. Our work provides a systematic way to catalog these possibilities.

Finally, we place our work in the context of related work in superconducting quantum circuit design. Early approaches generally optimized the parameters of a prescribed circuit. For example, \citet{YanSungOliver:2020} consider a generalized flux qubit and analytically explore the design space using approximations. While their analysis captures many previous single-loop qubit designs, it does not straightforwardly apply to general multi-loop, multimode circuits. In Ref.~\cite{LiuWangPan:2023}, Liu et al. use a variational quantum algorithm to optimize the component values of a parallel \texttt{JCL} circuit, where the cost function is a weighted sum of the qubit frequency and anharmonicity.

More recent work has incorporated circuit structure into the design problem by integrating machine learning and other advanced optimization methods into circuit design. \citet{Menke:2021aa} use a closed-loop swarm optimization algorithm to optimize desired circuit properties, given a circuit topology. \citet{CardLopeSanz2023} employ genetic algorithms to design noise-resilient qubits with particular selection rules. In related work by some of the same authors, this algorithm is used to design a new four-node qubit called the ``Difluxmon''~\cite{garcia-azorin_2024}. \citet{GraphQ_PrincetonMM_2024} use graph-based machine learning methods to address the circuit synthesis problem. These methods demonstrate the value of searching over circuit structure, but they generally explore restricted or heuristically generated subsets of the possible circuit space.

Parallel research has developed increasingly powerful methods for optimizing circuits once a topology has been selected, including gradient-based and differentiable optimization of circuit parameters \cite{rajabzadeh_2024, Wang2025SuperGrad}, circuit decomposition and classification \cite{smitham2024reciprocallumpedelementsuperconductingcircuits}, and automated optimization or inverse design of physical device geometries \cite{Shanto2024SQuADDS,Eriksson2025QDesignOptimizer,Seidel2026InverseDesign,Peng2026SQChipEM}. Recent work has also approached the inverse design directly at the Hamiltonian level by constructing circuits that approximate a desired nonlinear potential \cite{Hajihosseini2025Trainmon}.

Circuit enumeration addresses a distinct and complementary part of this inverse-design problem. Previous work has also considered the uniqueness and organization of superconducting circuit design spaces~\cite{rajabzadeh_2024,smitham2024reciprocallumpedelementsuperconductingcircuits}, but generally without attempting a complete enumeration. Our approach seeks to catalog the design space itself in the style of Foster and Campbell: we construct the complete set of unique circuits with a specified number of nodes and organize them into Hamiltonian classes. The resulting database provides a discrete search space that can be combined with existing optimization methods, giving future circuit synthesis efforts a concrete starting point for identifying and comparing candidate designs.

\section{Circuit Enumeration Algorithm}\label{sec:CirEnumAlg}

\begin{figure*}[ht]
\centering 
\includegraphics[width=0.99\textwidth]{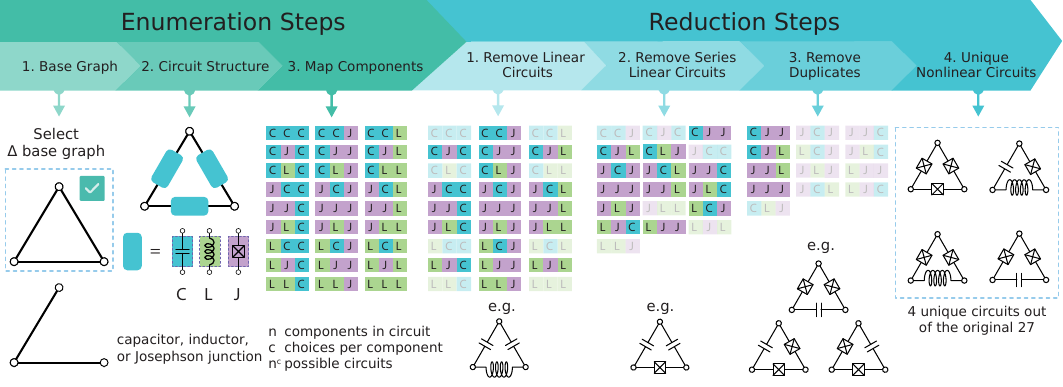}
\caption{Example circuit enumeration and reduction for three-node circuits. {\bf Enumeration:} (Step 1) For a given number of nodes, select a base graph. For $n=3$ there are two options, and for this figure, we choose the Delta topology. (Step 2) Given the Delta topology, we allow each edge to be one of $m$ components. In this example, the edge can be either a capacitor, inductor, or Josephson junction. In general, we also allow combinations of the three basic elements so the edges could be in the set \{\texttt{C}, \texttt{J}, \texttt{L}, \texttt{CL}, \texttt{JL}, \texttt{JC}, \texttt{JCL}\}. (Step 3) We enumerate all possible $m$-ary strings. In this example, there are 27 possible configurations. {\bf Reduction:} Circuits that are eliminated at each step are grayed out. (Step 1) We remove any circuit that does not contain a \texttt{J}. (Step 2) We remove circuits that have linear components in series because there is an equivalent circuit with one fewer node. (Step 3) Of the remaining circuits, several are equivalent, so we perform a graph isomorphism check to remove equivalent circuits. (Step 4) Four unique nonlinear circuits remain out of the original 27.}
\label{fig:sample_reduction}
\end{figure*}

Our enumeration procedure consists of two main steps:
\begin{enumerate}
    \item {\em Combinatorial enumeration.} \\ We represent circuit topologies using a set of unique undirected graphs. Each edge represents a location for a two-terminal circuit component, like a capacitor. For every topology, we enumerate all possible assignments of components to these edges.
    \item {\em Reduction of equivalent configurations.} \\ The combinatorial set generated in step 1 contains many identical circuits. We construct a minimal set of ``unique'' circuits. 
\end{enumerate}
The remainder of this section details these two steps, with a graphical summary provided in \cref{fig:sample_reduction}. Our procedure builds on the work of Herber et al. \cite{Herber:2017aa, herber:2017ab, herber_enhancements_2020}.

\begin{figure}[ht]
\centering 
\includegraphics[width=0.99\columnwidth]{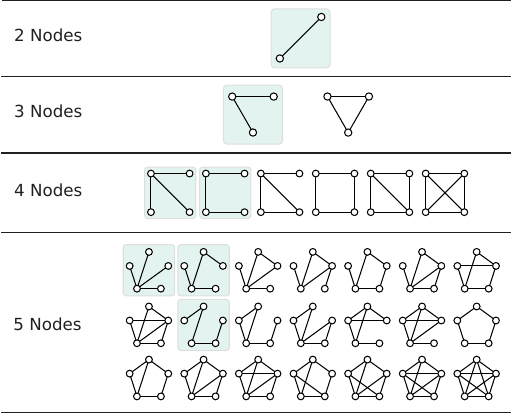}
\caption{The set of ``base graphs,'' or non-isomorphic, connected, undirected graphs with the specified number of nodes. Each represents a unique circuit topology.  All except for the fully connected five-node graph can be drawn as a planar graph. Acyclic graphs, or ``trees,'' are highlighted in blue.}\label{fig:basegraph_grid}
\end{figure}

\subsection{Combinatorial Enumeration}
\label{sec:comb_enum_subsec}

Our combinatorial enumeration generates all circuits with up to $n$ nodes built from linear capacitors (\verb|C|), linear inductors (\verb|L|), and Josephson junctions (\verb|J|). We choose \verb|C|, \verb|J|, and \verb|L| because most nonlinear superconducting circuits are built exclusively from these three elements. Throughout this paper, we will assume that Josephson junctions always have an accompanying intrinsic capacitance. While not strictly necessary for the combinatorial enumeration, this assumption is physically accurate and simplifies the analysis of multimode circuits. In superconducting-circuit diagrams, Josephson junctions with the junction capacitance omitted or shown explicitly are conventionally denoted, respectively, by
\begin{equation}
\begin{tikzpicture}[line width=0.8pt,baseline=(current bounding box.center),scale=0.7]
\pic at (0,1.1) {jj};
\node[anchor=west] at (1.25,1) {\small no junction capacitance,};
\pic at (0,0) {jjbox};
\node[anchor=west] at (1.25,0) {\small junction capacitance.};
\end{tikzpicture}
\end{equation}
We only use the latter in this work.

We define unique circuit topologies by the minimal set of connected, undirected, nonisomorphic graphs, which we obtain from Brendan McKay's combinatorial database \cite{McKay_Combinitorial_Data}. We refer to these graphs as ``base graphs'' and show them in \cref{fig:basegraph_grid}. There are 1, 2, 6, and 21 such topologies for circuits with two, three, four, and five nodes, respectively. The count rises to 112 for six nodes, 853 for seven nodes, and 11,117 for eight nodes.

Next, we generate all possible assignments of circuit elements to the edges of each base graph. For a base graph with $N_E$ edges and $m$ two-terminal components, there are $m^{N_E}$ such assignments, with each configuration naturally represented by a base-$m$ string of length $N_E$. Allowing for parallel combinations of the three fundamental elements, we define $m = 7$ allowed edge types. Their string representations are
\begin{equation}
\verb|C|,\, \verb|J|,\, \verb|L|,\, \verb|JC|,\, \verb|CL|,\, \verb|JL|,\, \textrm{and}\, \verb|JCL|.
\end{equation}

For example, three-node circuits utilize two base graphs (containing two and three edges, respectively), yielding $7^3 + 7^2 = 392$ total configurations. Similarly, summing the contributions across all base graphs for four and five nodes yields 139,944 and 338,332,113 raw assignments. Many of these configurations are equivalent; reducing them to a minimal set of ``unique'' circuits is discussed in \cref{sec:equiv_reduction}.

\subsection{Reduction of Equivalent Configurations}
\label{sec:equiv_reduction}
To construct a minimal set of ``unique'' nonlinear circuits, we remove configurations that are fully linear, contain reducible series linear components, or are otherwise topologically equivalent to another configuration. These three steps are illustrated in \cref{fig:sample_reduction}.

\subsubsection{Fully Linear Circuits}
Linear circuits are well studied, so our focus is only on nonlinear circuits. Although quantum information can be encoded in a linear circuit, like an LC resonator, nonlinear interactions are needed for universal gates or for encoding into a bosonic code~\cite{Ofek_binomial_2016,Ni_Binomail_2023,Sivak_GKP_2023}. We therefore discard any configuration that lacks a Josephson junction --- see \cref{fig:sample_reduction} step (1).

\subsubsection{Series Linear Components}
Two identical linear elements, such as capacitors or inductors, connected in series are equivalent at the outer terminals to a single element. A simple example appears in reduction step (2) of \cref{fig:sample_reduction}, where a circuit with two series capacitors is discarded because it is equivalent to the transmon circuit.

This extends naturally to $n$ linear elements of the same type meeting at a single node. They can be reduced to an equivalent network on the outer terminals alone. The case $n=3$ is the familiar star-mesh, or $\Delta$-Y, transformation \cite{kennelly_delta_wye_1899, rosen_new_1924}. In the Lagrangian formalism, these star configurations appear as non-dynamical modes. As shown in \cref{app:details:islands}, eliminating those modes from the circuit Lagrangian exactly recovers the generalized star-mesh transformation.

We therefore discard any circuit that contains one of these reducible star configurations, since its essential dynamics are already captured by a circuit with fewer nodes.

\subsubsection{Component-like Isomorphism}
Two circuit diagrams represent the same physical system if they differ only by a permutation of node labels. For example, in reduction step (3) of \cref{fig:sample_reduction}, the combinations \texttt{CJJ}, \texttt{JCJ}, and \texttt{JJC} all describe the same circuit because of the rotational symmetry of the underlying $\Delta$ base graph. We therefore retain one configuration, \texttt{CJJ}.

To identify these topological equivalences systematically, we test for isomorphism using a ``component graph'' representation~\cite{Herber:2017aa}. In a component graph, each circuit element and each circuit node is mapped to a colored vertex. Elements of the same type share a color, while nodes are colored by the number of elements connected to them. If two circuits are topologically equivalent, their component graphs are isomorphic \cite{Herber:2017aa, herber:2017ab, herber_enhancements_2020}.

\begin{figure}[hbt]
\centering 
\includegraphics[width=0.99\columnwidth]{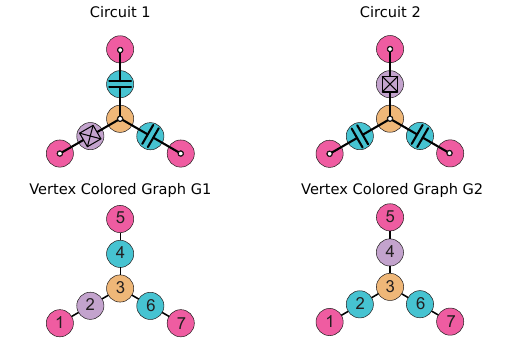}
\caption{Two equivalent circuits (top), along with their component graph representations (bottom). Each circuit component and node in the circuit diagram is represented by a colored vertex. Different circuit components are assigned different colors (capacitor $\rightarrow$ blue, Josephson junction $\rightarrow$ purple). Each node in the circuit is assigned a color based on its degree (1 connection $\rightarrow$ pink, 3 connections $\rightarrow$ orange). An isomorphism exists from $G_1 \rightarrow G_2$ with vertex mapping $\{1 \rightarrow 5, 2 \rightarrow 4, 3 \rightarrow 3, 4 \rightarrow 2, 5 \rightarrow 1, 6 \rightarrow 6, 7 \rightarrow 7\}$.} 
\label{fig:obvious_isomorphism}
\end{figure}

While graph isomorphism can be computationally expensive in general, the small graphs considered here make this approach practical. As illustrated in \cref{fig:obvious_isomorphism}, isomorphism between component graphs provides a mathematical description of the underlying circuit symmetries.

We perform pairwise isomorphism checks with the Python package \texttt{NetworkX}~\cite{hagberg2008exploring_networkx}. To reduce the search cost, we compare only circuits with the same base graph and the same component counts, for example, 2 \texttt{C}, 1 \texttt{L}, and 1 \texttt{J}. \Cref{app:psuedo} gives pseudocode for constructing the resulting nonisomorphic set.

\section{Categorization by Hamiltonian Class}\label{sec:HamEnum}

We want to study the quantum properties of our enumerated circuits, so we need a Hamiltonian for each circuit. We follow the standard procedure: begin with a circuit Lagrangian and obtain the Hamiltonian using a Legendre transform \cite{devoret1995,VoolDevoret2017,Rasmussen2021}. Crucially, this procedure requires a choice of coordinates.

The circuit Lagrangian is most straightforwardly expressed in terms of node flux variables $\Phi = (\phi_1, \phi_2, \ldots,\phi_n )^T$:
\begin{align}\label{eq:node_lagrangian_paper}
    \mathcal{L}(\Phi, \dot \Phi) = \frac{1}{2}\dot\Phi^TC\dot\Phi &- \frac{1}{2}\Phi^TL^*\Phi  - E_{JJ}(\Phi) \, ,
\end{align}
where $\dot{\Phi}$ is the time derivative of $\Phi$, $C$ is the capacitance matrix, $L^*$ is the inverse inductance (susceptance) matrix, and $E_{JJ}(\Phi)$ is the Josephson energy. 
When defining Hamiltonian classes, we treat charge offsets and external fluxes as operating-point parameters. Including their continuous values would turn the finite structural classification into a continuous one. We therefore set them to zero for the classification and reintroduce them in the numerical analysis.

While node-flux variables provide a straightforward basis for writing the Lagrangian, they are rarely the best choice for understanding the dynamics.
To understand why, we can compare a grounded and an ideal floating transmon.
A grounded transmon has one non-ground node flux, which directly gives the phase difference across the Josephson junction.
A floating transmon is instead described by two node fluxes, $\phi_1$ and $\phi_2$, even though its internal dynamics depend only on their difference, $\theta =\phi_1-\phi_2$.
Introducing the common coordinate $\phi_+=(\phi_1+\phi_2)/2$ separates the internal transmon coordinate from collective motion of the two islands.
Thus, even in this simple example, changing coordinates isolates the physically relevant degree of freedom. Below, we apply this idea systematically to separate mode types and minimize intermode coupling, which provides the coordinates used to define Hamiltonian classes.

In general, we can define a new set of coordinates $\tilde{\Phi}$ via a linear transformation of node flux variables, $\tilde{\Phi} = Z^{-1}\Phi$. Substituting $\Phi = Z\tilde{\Phi}$ into \cref{eq:node_lagrangian_paper} gives the Lagrangian in the new coordinates.
\begin{align}\label{eq:tilde_lagrangian_paper}
    \mathcal{L}(\tilde{\Phi},\dot{\tilde{\Phi}})
    &=
    \frac{1}{2}\dot{\tilde{\Phi}}^T \tilde{C} \dot{\tilde{\Phi}}
    -
    \frac{1}{2}\tilde{\Phi}^T \tilde{L}^* \tilde{\Phi}
    -
    E_{JJ}(Z\tilde{\Phi})\,.
\end{align}
The quadratic terms are described by the transformed matrices $\tilde{C} = Z^T C Z$ and $\tilde{L}^* = Z^T L^* Z$. This Lagrangian can then be transformed to a Hamiltonian via the Legendre transform.

A single Hamiltonian can describe different operating regimes of the same circuit. For example, the Hamiltonian $E_C\, n^2 - E_J \cos \theta$ describes both the transmon and the Cooper-pair box for different ratios $E_J/E_C$.

\begin{figure}[bht]
\centering
\includegraphics[width=\columnwidth]{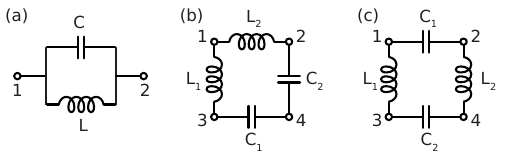}
\caption{Three circuits with the same Hamiltonian class under the coordinates $\varphi_a = \phi_2-\phi_1$, $\varphi_b = \phi_2 - \phi_3$, and $\varphi_c =  \phi_4-\phi_3 + \phi_2-\phi_1$. With $C=C_1C_2/(C_1+C_2)$ and $L=L_1+L_2$, they also have identical equations of motion.}
\label{fig:LC_three_circuits}
\end{figure}

Distinct circuits, however, can also share the same Hamiltonian structure.
Consider the topologically distinct circuits in \cref{fig:LC_three_circuits}, which reduce to the same Hamiltonian. With an appropriate choice of coordinates, each circuit has one dynamical degree of freedom, a kinetic term proportional to $q^2$, and a potential term proportional to $\varphi^2$. When circuits share this structure, relative to a specific choice of ``\canonical''~coordinates, we say that they belong to the same ``Hamiltonian class''.

\begin{definition}[Hamiltonian Class]\label{def:ham_class}
A set of Hamiltonians that share the same functional form, up to parameter values.
\end{definition}

Consequently, we denote Hamiltonian classes without parameters, e.g., for the transmon the class is $n^2 - \cos \theta$.
The Hamiltonian class of a circuit depends on the choice of ``\canonical~coordinates''.
\begin{definition}[\Canonical~Coordinates]\label{def:ref_coord}
For a set of circuits, \canonical~coordinates are a choice of valid coordinates for every circuit in the set. 
\end{definition}
The \canonical~coordinates thus establish a mapping from circuits to Hamiltonian classes. Different coordinate constructions can merge or split classes, and thus yield fewer or more classes.

Hamiltonian classes can aid circuit design by providing a bias-independent classification of mode and coupling structure. We present two applications. First, they enable an inverse lookup from a desired Hamiltonian structure to the circuit topologies that realize it. Circuits within the same class can nevertheless access different parameter regimes and present different implementation tradeoffs, as explored in \cref{sec:lookup}. Second, as shown in \cref{sec:delocalized}, the classification reveals topology-level constraints that can force genuinely collective, delocalized nonlinear coordinates.

\begin{figure}
\centering 
\includegraphics[width=\columnwidth]{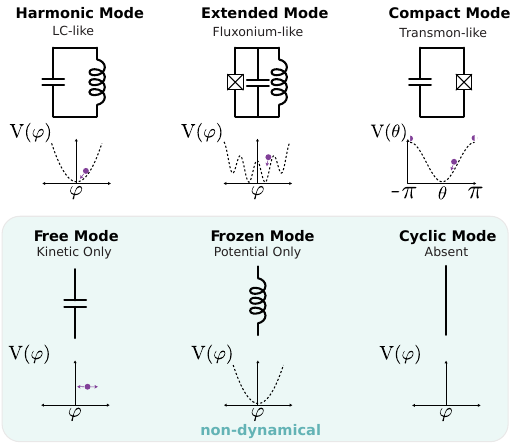}
\caption{Types of dynamical (top) and non-dynamical (bottom) modes present in superconducting circuits of pure capacitors, pure inductors, and Josephson junctions with intrinsic capacitance. Cartoons of potential energy are shown for each mode type, with a corresponding ``particle'' moving within the potential. Here and throughout we will denote harmonic and extended modes using $q,\varphi$ and compact modes using $n, \theta$. The compact mode has periodic boundary conditions, shown by the particle split between $\theta=-\pi$ and $\theta=\pi$.}
\label{fig:variable_kinds}
\end{figure}

In order to categorize circuits into Hamiltonian classes, we introduce an algorithm to generate \canonical~coordinates. We require it to return the same coordinates for circuits that differ only by a relabeling of nodes, that is, for circuits with isomorphic component graphs. Otherwise, Hamiltonian classes would depend on which circuit diagram is chosen to represent a given unique circuit. Additionally, we choose to consider only coordinates that are independent of circuit parameters. This ensures each unique circuit has a single choice of coordinates. The procedure consists of three steps, described in more detail in \cref{app:var_trans}:

\begin{enumerate}
    \item {\em Mode separation.} \\  For a circuit containing capacitors, inductors, and Josephson junctions (with intrinsic capacitance\footnote{We discuss the open question of treating circuits without intrinsic junction capacitance in \cref{app:var_trans}.}), we separate coordinates into six mode types, shown in \cref{fig:variable_kinds}.
    Compact variables (those with a periodic potential) are required to form an integer lattice with charge quantized in units of Cooper pairs.
    \item {\em Non-dynamical mode removal.} \\ We eliminate coordinates corresponding to constraints or redundancies, yielding an invertible capacitance matrix and a minimal set of dynamical variables.
    \item {\em Coupling minimization.} \\ Within the constraints of the previous two steps, we minimize intermode coupling, while keeping dynamical coordinates independent of circuit parameters.\footnote{Allowing parameter-dependent transformations can eliminate linear coupling at the cost of increased nonlinear coupling \cite{Rajabzadeh2023analysisofarbitrary}.} We define ``minimizing coupling'' as maximizing the number of strictly zero intermode coupling terms, first for nonlinear terms and then for linear terms.
    We compute a provably optimal solution using a custom solver built on \textsc{Singular} \cite{Singular_2024} and the \textsc{Z3} SMT solver \cite{DeMouraBjornerZ3_2008}. This step can be viewed as a linear change of coordinates within each dynamical subspace.
\end{enumerate}

\begin{figure}[t!h]
\centering 
\includegraphics[width=0.99\columnwidth]{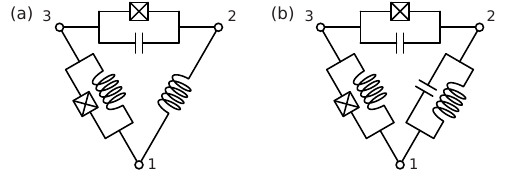}
\caption{Two circuits from classes (a) 3-11 and (b) 3-12 in \cref{tab:comparison_by_var} that have equivalent mode structures if a different choice of coordinates is used.}
\label{fig:diff_class_example}
\end{figure}

We now illustrate how the reference-coordinate selection minimizes coupling. For the two circuits in \cref{fig:diff_class_example}, the algorithm produces $\varphi_1 = \phi_1 - \phi_3$ and $\varphi_2 = \phi_2 - \phi_3$, where $\phi_i$ is the flux variable at node $i$. The resulting Hamiltonian classes for circuits (a) and (b) are
\begin{subequations}
\begin{align}
    q_{1}^{2} +\phantom{q_{1} q_{2}+} &q_{2}^{2} + \varphi_{1}^{2} + \varphi_{1} \varphi_{2} + \varphi_{2}^{2} + \cos{\varphi_{1} } + \cos{\varphi_{2}} \\
    q_{1}^{2} + q_{1} q_{2} + &q_{2}^{2} + \varphi_{1}^{2} + \varphi_{1} \varphi_{2} + \varphi_{2}^{2} + \cos{\varphi_{1} } + \cos{\varphi_{2}}.
\end{align}
\end{subequations}
The extra coupling in circuit (b) comes directly from the extra capacitor between nodes 1 and 2. If we had instead chosen $\varphi_2 = \phi_2 - \phi_1$, both circuits would fall into the same Hamiltonian class
\begin{align} \label{eq:extraHform}
{q}_{1}^{2} +& {q}_{1} {q}_{2} + {q}_{2}^{2} + {\varphi}_{1}^{2} + {\varphi}_{2}^{2} + \cos{\left({\varphi}_{1} - {\varphi}_{2} \right)} +  \cos{{\varphi}_{1}} \, .
\end{align}
This alternative choice replaces linear coupling, ${\varphi_1}{\varphi_2}$, with nonlinear coupling, $\cos({\varphi_1}-{\varphi_2})$. Our procedure prioritizes minimizing nonlinear coupling and then, within that constraint, minimizing linear coupling.

\section{Enumeration Results}\label{sec:results}

\Cref{tab:overview} summarizes our enumeration results by number of nodes. The second column shows a combinatorial explosion of the naive circuit count from \cref{sec:comb_enum_subsec}. The third column counts unique circuits using methods from \cref{sec:equiv_reduction}, showing that the numbers are far smaller than naive counting suggests. We show similar summary statistics in \cref{app:additional_results} as a function of the number of junctions, compact variables, and phase loops in a circuit.

\begin{table}[h!]
\setlength{\tabcolsep}{13pt}
\begin{tabular}{@{}cccc@{}}
\toprule
Nodes      &  Circuits&  Unique Circuits  &  Classes \\ \toprule
2                  & 7          &   4                  & 2  \\ \midrule
3                  & 392        &   80                 &  19 \\ \midrule
4                  & 139,944    &   10,982             &  232 \\\midrule
5             & 338,332,113  &  7,976,949          &  4893 \\\bottomrule
\end{tabular}
    \caption{
    Number of circuits and Hamiltonian classes for circuits with two to five nodes. The first column gives the number of nodes. The second gives the naive circuit count, $7^{N_E}$, summed over all base graphs for that node number (see \cref{fig:basegraph_grid}). For three-node circuits, this gives $7^3 + 7^2 = 392$ from the two base graphs. The third column gives the number of unique circuits containing at least one Josephson junction. Because the naive count includes many equivalent circuits, the reduction from column 2 to 3 is large. The fourth column gives the number of Hamiltonian classes, namely distinct combinations of mode types and inter-mode couplings identified from the unique circuits.}\label{tab:overview}
\end{table}

This combinatorial enumeration defines the circuit design space. However, unique circuits can behave similarly. In the fourth column of \cref{tab:overview}, we use the transformation developed in \cref{sec:HamEnum} to group circuits into Hamiltonian classes that capture these shared behaviors. Recall that the number of Hamiltonian classes depends on the choice of \canonical\ coordinates, so the counts reported here correspond to our specific construction.
For circuits with up to five nodes, this technique reduces millions of unique circuits to fewer than five thousand Hamiltonian classes. A full database of these circuits is available through \cite{weissler_2026_zenodo_data}.

\begin{table*}[ht!]
\includegraphics[width=0.99\textwidth]{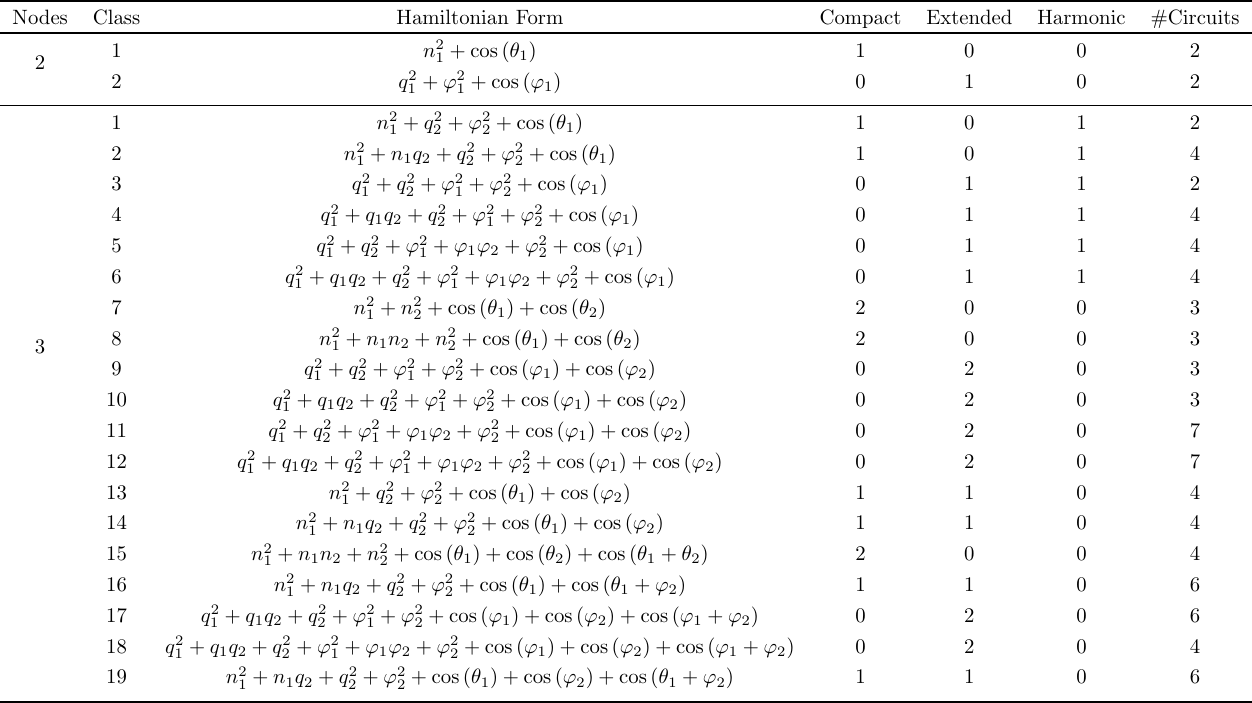}\\[0.0cm]
    \includegraphics[width=0.99\textwidth]{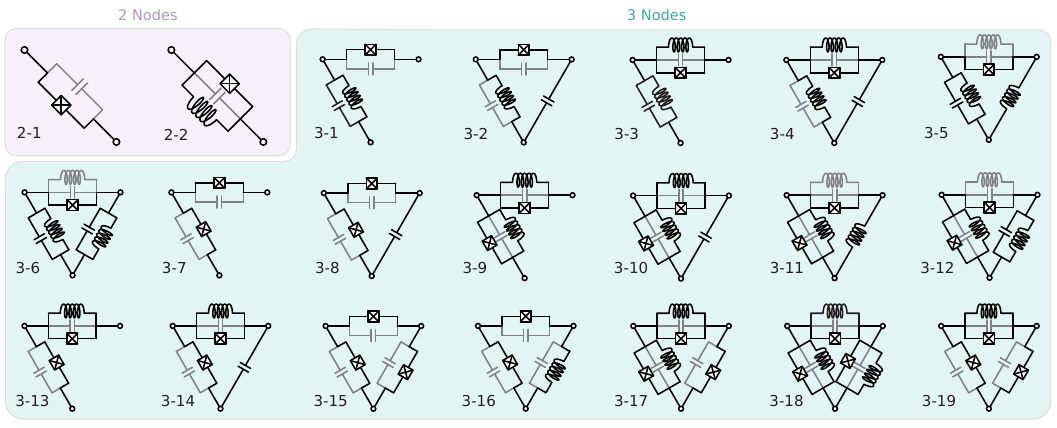}\\[0.0cm]
    \vspace{0.1cm}
    \hrule
    \vspace{0.1cm}
    \caption{Hamiltonian classes (top) and corresponding circuits (bottom) up to 3 nodes in size. Compact variables are denoted by $n, \theta$, while extended and harmonic variables are denoted by $q, \varphi$. Grayed-out circuit elements may be included or excluded without changing the class. Because each Josephson junction includes a small intrinsic capacitance, most optional elements correspond to capacitors placed in parallel with junctions. Optional inductors, such as those in classes $3-5$, $3-6$, $3-11$, and $3-12$, are present when the circuit has no compact modes and already contains a series inductor across the specified branch. Likewise, the optional capacitors in classes $3-2$, $3-4$, and $3-16$ occur when that branch already contains a series capacitor.}
    \label{tab:comparison_by_var} 
\end{table*}

In \cref{tab:comparison_by_var}, we list the Hamiltonian classes for two- and three-node circuits.\footnote{A similar table of Hamiltonian classes for four- and five-node circuits is available on our repository \cite{sircuitenum_repo}.} Each class is labeled by its node number and class number, $[\rm Nodes]-[\rm Class]$. The ``Hamiltonian Class'' column gives the Hamiltonian with coefficients omitted, since the coefficients vary across circuits. The ``Compact,'' ``Extended,'' and ``Harmonic'' columns list the mode types explicitly, although they can also be inferred from the Hamiltonian.

Two-node circuits are either fluxonium-like, with one extended mode, or transmon-like, with one compact mode. Three-node circuits fall into 19 classes. Most of them, classes 3-1 through 3-14, can be understood as individual or linearly coupled transmon-like, fluxonium-like, or LC-like systems. The remaining classes exhibit nonlinear coupling. They include all circuits with three Josephson junctions, as well as the two-junction Bifluxon circuit 3-16 \cite{kalashnikov_bifluxon_2020}.

The ``\# Circuits'' column gives the number of circuits in each class. The corresponding diagrams appear in the bottom panel, with gray elements indicating optional elements that do not change the class. Because we include a small intrinsic capacitance for every Josephson junction, most optional elements are capacitors in parallel with junctions. The optional inductors in classes $3-5$, $3-6$, $3-11$, and $3-12$ arise when the circuit has no compact modes and already contains a series inductance across the specified branch. The optional capacitors in classes $3-2$, $3-4$, and $3-16$ similarly arise when there is already a series capacitance across the specified branch.

The three-node circuits in \cref{tab:comparison_by_var} suggest three design principles. First, the low-node design space is far smaller than raw combinatorics suggests. Many apparently different circuit diagrams collapse to the same behavior. Second, the types of modes present are determined by the inductive portion of each circuit (e.g., \texttt{J}, \texttt{L}). In three-node circuits, capacitors only modify the ``masses'' of the individual modes and alter the intermode charge coupling, although for larger circuits they can also convert frozen modes into harmonic modes. Finally, most classes can be viewed as coupled transmon-like, fluxonium-like, or LC-like modes. Only one three-node class, $3-16$, features a Hamiltonian that cannot be straightforwardly decomposed into these component modes, with respect to our reference coordinates. 

\subsection{Inverse Lookup: Find all Circuits in a Class}\label{sec:lookup}

As an application of Hamiltonian classification, we consider an inverse-lookup problem. Given an existing circuit (or equivalently a target Hamiltonian), find all circuits within the same Hamiltonian class. We consider two four-node examples, the Difluxmon circuit \cite{garcia-azorin_2024} and the \zp\ circuit \cite{Gyenis_zero_pi_2021a}, which is a protected qubit. The Difluxmon class has similar membership rules to the three-node circuits, while \zp\ is an example of a system that exhibits emergent dynamics when imposing circuit symmetry.

\begin{figure}[th]
\centering 
\includegraphics[width=0.99\columnwidth]{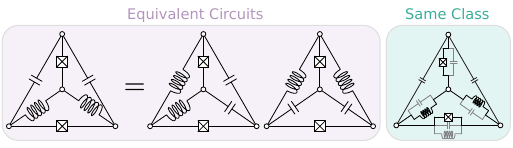}
\caption{(Left) Topologically equivalent circuits to the Difluxmon circuit \cite{garcia-azorin_2024}, up to trivial rotations and reflections. (Right) All circuits in the same Hamiltonian class as the original Difluxmon circuit. Gray elements can be added or removed without changing the Hamiltonian class and represent additional capacitances and inductances between nodes with existing capacitive or inductive paths.} \label{fig:difluxmon_isomorphism}
\end{figure}

In \cref{fig:difluxmon_isomorphism}, we show all circuits in the same Hamiltonian class as the Difluxmon circuit. Our framework identifies two distinct inductive circuits, with many possible capacitor placements. As with the three-node Hamiltonian classes, the optional elements represent capacitances and inductances across branches that already have a capacitive or inductive path, respectively. In this sense, the Difluxmon can be viewed as a specific instantiation within the specified Hamiltonian class, with deliberately small capacitances on all but two circuit branches.

\begin{figure}[ht!]
\centering
\includegraphics[width=0.99\columnwidth]{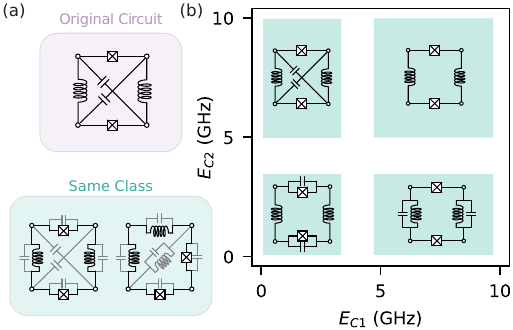}
\caption{(a) All circuits in the same Hamiltonian class as the original \zp\ circuit. Gray capacitors must be included or excluded in pairs across identical elements to preserve the required symmetry. For the second class, which contains the Bifluxon circuit \cite{kalashnikov_bifluxon_2020}, the series inductors may be excluded if the diagonal inductor is included. (b) Parameter regions accessible to four different capacitive layouts of the class containing the original \zp\ circuit, assuming junction charging energies of 5--10 GHz and capacitor charging energies of 0.05--5 GHz. Only the original \zp\ circuit (top left) achieves the small $E_{C1}$ and large $E_{C2}$ required for protection. 
The top right circuit has no dedicated capacitors, but both modes still have charging energy because we include the junction's intrinsic capacitance. The Bifluxon-class circuits, not depicted in (b), can achieve every combination of $E_{C1}$ and $E_{C2}$, except for the protected small $E_{C1}$ and large $E_{C2}$ case.}\label{fig:copycat_qubits}
\end{figure}

We now turn our attention to the \zp\ circuit, which has the same base graph as the Difluxmon, but with one inductor and one capacitor interchanged. Assuming identical pairs of capacitors, inductors, and junctions, its Hamiltonian is
\begin{equation}\label{eq:zp_ham}
    H_{0-\pi} = E_{C1} n_1^2 + E_{C2} q_2^2 + E_L \varphi_2^2 - 2E_J \cos \theta_1 \cos \varphi_2 \, ,
\end{equation}
ignoring a completely decoupled harmonic mode. This system is intrinsically protected against decay and dephasing when $E_{C1}/E_J$ is small and $E_{C2}/E_L$ is large \cite{paolo_control_2019}. In \cref{fig:copycat_qubits}, we examine all three- and four-node circuits that belong to this Hamiltonian class.

The standard \zp\ circuit is shown in \cref{fig:copycat_qubits}(a). Through our enumeration, we find two distinct inductive circuits in the same Hamiltonian class. Each inductive circuit can have many possible capacitor placements. The second inductive circuit corresponds to the Bifluxon circuit \cite{kalashnikov_bifluxon_2020}. While these circuits share the same mode structure, their additional capacitors set drastically different effective parameters $E_{C1}$ and $E_{C2}$. Moreover, depending on the capacitive configuration, an additional harmonic mode can also emerge, as in the original \zp\ circuit. As shown in \cref{fig:copycat_qubits}(b), only the original \zp\ circuit reaches the regime of small $E_{C1}$ and large $E_{C2}$ required for protection.

This example also highlights a broader methodological point. Many proposed multimode protected qubits, including the \zp\ circuit, rely on highly symmetric parameter regimes in which pairs of capacitances, inductances, and Josephson energies are equal, causing coupling terms that depend on the difference of those parameters to go to zero  \cite{Gyenis_zero_pi_2021a, kalashnikov_bifluxon_2020, kitaev_protected_2006, smith2020TwoCooperPairTunneling}. In the general enumeration, each circuit element is treated as an independent symbolic parameter. Here, however, we imposed exact symmetry between matching capacitors, inductors, and junctions to capture this decoupling. Our coordinate-generation algorithm is capable of identifying these special regimes where symmetric parameters simplify the dynamics by decoupling certain modes.

\subsection{Identification of Delocalized Coordinates}\label{sec:delocalized}

\begin{figure}[t]
    \centering
    \includegraphics[width=0.99\linewidth]{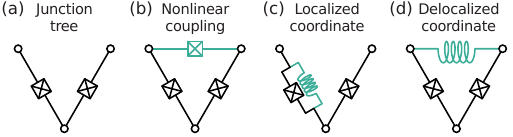}
    \caption{Topological origins of localized and delocalized extended coordinates. (a) The three-node Josephson junction tree and (b--d) three possible single-branch inductive modifications, with each added branch shown in teal. (b) A loop-completing junction nonlinearly couples two compact coordinates. (c) An inductive path in parallel with a single junction converts one compact coordinate into an extended coordinate that can remain localized to that junction. (d) An inductor completing a mixed loop instead creates an extended coordinate that must be delocalized across both junctions. This is the inductive skeleton of Hamiltonian class $3-16$.}
    \label{fig:ind_cir_three}
\end{figure}

\begin{figure*}[bht]
\centering
\includegraphics[width=\textwidth]{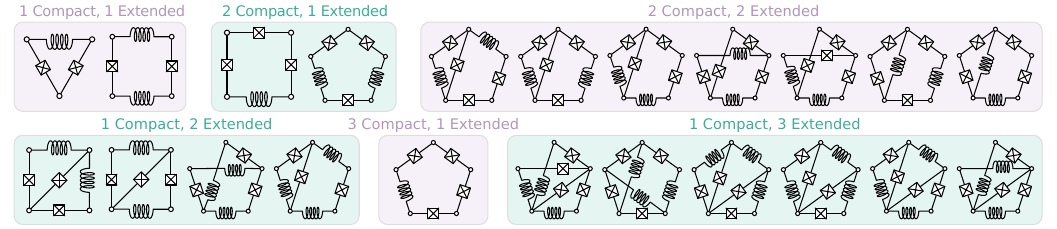}
\caption{Catalog of irreducible inductive skeletons up to five nodes in size with at least one necessarily delocalized extended coordinate. No displayed junction is shunted by an inductor-only path, so all extended coordinates are delocalized (see \cref{app:skeleton}). The skeletons are separated by their numbers of compact and extended modes. There is one three-node skeleton, four four-node skeletons, and seventeen five-node skeletons. We omit skeletons containing shunted junctions, nodes connected only to inductors, or all-junction or all-inductor loops. We also omit circuits that become disconnected when one node is removed. Extra capacitors may create harmonic modes and couplings but do not modify the nonlinear coordinate structure.}
\label{fig:nontrivial_classes}
\end{figure*}

The Hamiltonian classes provide a compressed catalog of the enumerated circuits, but they can reveal more than redundancy. 
Some circuit topologies force an extended coordinate to be \emph{delocalized} over several circuit components.
Remarkably, this can be identified directly from the \emph{inductive skeleton}---the subcircuit formed only by junctions and inductors---before any capacitances or component values are chosen.

We call an extended nonlinear coordinate \emph{localized} if it can be chosen as the phase difference across a single Josephson junction. The three-node circuits in \cref{fig:ind_cir_three} illustrate when this is and is not possible.
Starting from the two-junction tree in (a), adding a junction closes a junction-only loop and couples the two compact coordinates, as shown in (b). Adding an inductor in parallel with one junction instead converts one compact coordinate into an extended coordinate that remains localized to that junction,
as in (c). In contrast, closing a loop containing two junctions and an inductor, as in (d), produces an extended coordinate that necessarily involves both
junctions. Hence the coordinate is delocalized. This is the inductive skeleton of Hamiltonian class $3-16$, which contains the Bifluxon circuit~\cite{kalashnikov_bifluxon_2020}.

The same loop structure gives a general criterion. A necessarily delocalized coordinate requires a mixed loop containing multiple junctions and at least one inductor. If no individual junction is shunted by an inductor-only path, such a loop is also sufficient.
 In \cref{app:skeleton}, we derive the general criterion for any circuit to have a delocalized extended nonlinear coordinate (or coordinates). We give a procedure to determine the exact count $\delta$ for a circuit with any number of nodes. Importantly, $\delta$ depends only on the inductive skeleton.

After removing inductively shunted junctions and other redundancies (see \cref{app:skeleton}), we obtain one three-node, four four-node,
and seventeen five-node irreducible-skeletons with necessarily delocalized
coordinates. These are circuits where all extended modes are delocalized, and they are shown in \cref{fig:nontrivial_classes}. Thus the Hamiltonian categorization can highlight circuits with collective nonlinear degrees of freedom without constructing and diagonalizing every capacitive realization.

Delocalization does not by itself imply noise protection. 
We have already seen that the Bifluxon is a three-node example of a protected qubit with coordinate delocalization. 
It turns out that the \zp\ circuit is a four-node example as well.  
In symmetric coordinates, its Josephson potential is $-2E_J\cos\theta_1\cos\varphi_2$. Equivalently, the two physical junction phases involve the combinations $\theta_1\pm\varphi_2$.
Its inductive skeleton is shown in \cref{fig:nontrivial_classes} and \cref{fig:copycat_qubits}. This collective nonlinear structure is one reason that delocalized skeletons provide a natural subset of circuits to investigate when searching for protected qubits.

\section{Searching For New Qubits}\label{sec:optimization}

The main use for our superconducting circuit enumeration is to enhance circuit synthesis methods. As a proof of concept, we search for new three-node qubit circuits. 
The search performed here uses the enumerated set of unique circuits directly; it does not use Hamiltonian classes to reduce the optimization space. A class-level optimization could substantially reduce this search space, but a Hamiltonian class does not uniquely determine either the physically accessible parameter region or circuit-specific noise couplings. We therefore optimize explicit circuit realizations.

We begin by developing a framework to optimize the performance of an arbitrary circuit as a qubit and then apply it to each unique three-node circuit. Any circuit can be used as a superconducting qubit, but some perform better than others. We quantify this performance as the number of gates $N_{\rm gate}$ that can be performed within the coherence time $T_2$ 
\begin{align}\label{eq:ngate}
N_{\rm gate} = \frac{T_2}{T_{\rm gate}}\,.
\end{align}
Here $T_{\rm gate}$ estimates the duration of a single qubit $X$ gate, and we define $1/T_2 = 1/2T_1 + 1/T_\phi$ to include both energy relaxation $T_1$ and dephasing $T_\phi$.

Because we consider many previously unstudied circuits, our modeling requires several simplifying assumptions, which we describe here and in \cref{app:details}. Our objective function $N_{\rm gate}$ is best viewed as a {\em qualitative} architecture-agnostic screening tool for rapid evaluation, not as a quantitative predictor of gate count. It highlights circuits and operating points that balance coherence and control.

To estimate $T_2$, we use the \verb|SQcircuit| package to analyze our circuits \cite{Rajabzadeh2023analysisofarbitrary, rajabzadeh_2024}. Energy-relaxation rates are calculated using Fermi's golden rule, while dephasing rates depend on the sensitivity of the qubit frequency to noise parameters. We use the default noise amplitudes, which agree with the lower bound presented in \cite{houck_life_2009}. For more details on how these rates are calculated, see \cite{Rajabzadeh2023analysisofarbitrary, rajabzadeh_2024}. Moreover, for circuits with external flux, \texttt{SQcircuit} assigns the branch fluxes using its gauge-consistent irrotational construction~\cite{YouKoch2019,BryonExtFlux2023,Rajabzadeh2023analysisofarbitrary}. Flux-noise estimates therefore use the resulting circuit-specific flux distribution.

To estimate $T_{\rm gate}$, we consider a simplified three-level system and estimate the time required to complete either a direct transition $\ket{0} \rightarrow \ket{1}$ or a transition via the higher-lying state $\ket{0} \rightarrow \ket{2} \rightarrow \ket{1}$. Our model, detailed in \cref{app:gate_time}, uses the level spacing to compute a lower bound on the gate time, independent of any specific drive.

While estimating gate time from the matrix elements of an explicit drive operator is more precise, constructing a physical drive model for every candidate is computationally prohibitive. Identifying and optimizing these circuit-specific couplings therefore lies beyond the scope of this enumeration study. We instead rely on our analytical speed limit for efficient optimization. Then, for the top identified circuit, we construct an explicit drive circuit and show that the simulated gate speed closely matches our analytical model.

We optimize \cref{eq:ngate} for each circuit; the optimization dimension is given by the number of circuit elements and offset biases.
We execute this search using a Python implementation of the differential evolution algorithm~\cite{price_benchmarking_2005}, a widely used gradient-free genetic algorithm. Our optimization returns the optimal circuit-element parameters and offset biases, which we call the ``optimized point''.
We also perform a sensitivity analysis to model imperfections in fabrication. Good candidate qubits are robust to small changes in the optimized parameters. We do this by randomly sampling 100 nearby points in circuit parameter space from a normal distribution with a 2.5\% standard deviation about the optimized point.

A key choice in the optimization is the allowed values for circuit parameters. We choose ranges that extend somewhat beyond present-day superconducting qubits (see \cref{tab:params_overview}). This choice reflects the possibility that design, materials, and fabrication advances may make such devices feasible. We note that, in the first-pass optimization over all circuits, every Josephson junction includes the fixed intrinsic capacitance $C_J=1.9\,{\rm fF}$.

\begin{table}[ht]
\setlength{\tabcolsep}{4.5pt}
\begin{tabular}{@{}crrcc@{}}
\toprule
Parameter \,\,      &  Min (Freq) & Max (Freq) & Min (SI) &  Max(SI) \\ \toprule
$E_C$,~ $C$                  & 0.05 GHz &  10.0 GHz & 1.9 fF  & 0.39 pF \\ \midrule
$E_L$,~ $L$                  & 0.05 GHz &  \phantom{1}5.0 GHz  & 33 nH  & 3.3 $\mu$H \\ \midrule
$E_J, ~L_J$                  & 1.0\phantom{0} GHz   & 30.0 GHz &  5.4 nH & 0.16 $\mu$H \\ \midrule
$E_{C_J}$, ~$C_J$ & 10.0\phantom{0} GHz & 10.0 GHz & 1.9 fF & 1.9 fF \\\bottomrule
\end{tabular}
\caption{Allowed ranges for circuit parameters. The junction capacitance $C_J$ is taken to be fixed and matches the smallest capacitance allowed. For details on how to convert between physical and frequency units see \cref{app:params}. }
\label{tab:params_overview}
\end{table}

To assess this method, we compare our model's predictions to devices from the literature in \cref{app:sweeps}. Accordingly, in \cref{app:sweeps} we restrict the circuit parameters to values that can be realized with current technology. We sweep the circuit parameter space for transmon and fluxonium-like circuits and find that high-performing transmon devices from the literature lie within the best performing regions predicted by our model. We find similarly good agreement in fluxonium if we slightly modify the noise model. Furthermore, our gate time estimate is able to predict the relative gate times of the considered devices.

\subsection{Two Node Circuits}
\begin{table}[ht]
\setlength{\tabcolsep}{5.0pt}
\centering
\begin{tabular}{ccccc}
\toprule
Base Graph & Circuit & Class & $N_{\rm gate}$ (Mean) & $N_{\rm gate}$ (Std) \\ \toprule
\multirow{4}{*}{\includegraphics[]{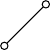}}& \texttt{J} & 2-1 & $3.75\cdot10^{3}$ & $1.8\cdot10^{2}$ \\
& \texttt{JL} & 2-2 & $8.99\cdot10^{4}$ & $1.4\cdot10^{3}$ \\
& \texttt{JC} & 2-1 & $4.14\cdot10^{4}$ & $8.6\cdot10^{2}$ \\
& \texttt{JCL} & 2-2 & $1.87\cdot10^{5}$ & $6.0\cdot10^{3}$ \\
\bottomrule
\end{tabular}
\caption{Optimized number of gates for all two-node circuits. The maximum value, $N_{\rm gate} = 187,000 \approx 1.9\times 10^5$, for the optimized fluxonium serves as a benchmark for larger circuits.} 
\label{tab:optim_two}
\end{table}

There are four unique two-node circuits. We optimize their performance over the parameter ranges given in \cref{tab:params_overview}. As shown in \cref{tab:optim_two}, the fluxonium-like \verb|JCL| circuit scores highest among the two-node circuits, with an average $N_{\rm gate} = 187,000$. This value is obtained for $E_C = 2.63$ GHz, $E_J = 2.85$ GHz, $E_L = 0.51$ GHz, and $\varphi_{\rm ext} = 0$. As we consider three-node circuits, $N_{\rm gate} = 187,000 \approx 1.9\times 10^5$ will serve as a benchmark.

\begin{figure}[ht]
\centering 
\includegraphics[width=0.99\columnwidth]{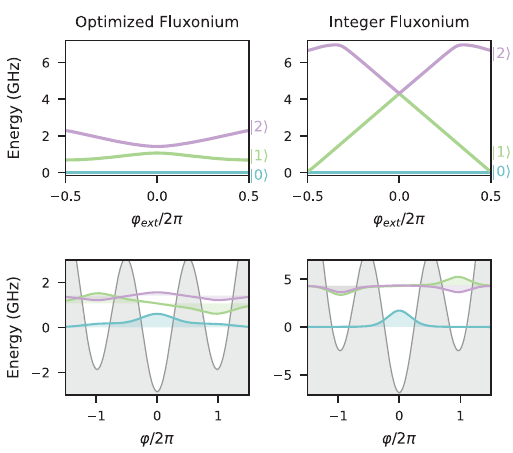}
\caption{Dependence of the first three energy levels on external flux (top), optimized fluxonium (left) and integer fluxonium (right), using parameters from \cite{mencia_integer_2024} device D. The first three eigenstates are plotted for both circuits (bottom), with global phase calibrated to make the peak value of the $\ket{0}$ state real and positive. The optimized fluxonium has $E_C = 2.63$ GHz, $E_J = 2.85$ GHz, and $E_L = 0.51$ GHz. The integer fluxonium has $E_C =$ 1.46 GHz, $E_J=$ 6.86 GHz, and $E_L$ = 0.23 GHz. Both circuits operate at zero external flux.}
\label{fig:fluxonium_diagram}
\end{figure}

In \cref{fig:fluxonium_diagram}, we show the energy spectrum and wave functions of the optimized fluxonium. It is similar to the ``Integer Fluxonium'' qubit. The ground state $\ket{0}$ lies mainly in the central well, while $\ket{1}$ and $\ket{2}$ lie mostly in a superposition of the wells at $\varphi = -\pi, \pi$ \cite{mencia_integer_2024}. Our optimized fluxonium acts as a ``light'' version of the integer fluxonium, with the large inductor flattening the flux dispersion. Neither system appears in the parameter sweep of \cref{app:sweeps}, because both operate at $\varphi_{\rm ext}=0$.

While the integer fluxonium is a compelling candidate, our optimization ultimately ranked it below our top-performing circuit for two reasons. First, integer fluxonium enjoys greater disjoint support than the optimized fluxonium, but it is more sensitive to external flux, which penalizes its performance within our noise model. Second, our spectral gate-time estimate penalizes circuits with nearly degenerate first and second excited states ($E_1 \approx E_2$), a defining feature of the integer fluxonium. This second penalty illustrates a key limitation of our gate-time model. In the integer fluxonium, the $\ket{0} \rightarrow \ket{2}$ transition is physically forbidden, meaning direct $\ket{0} \rightarrow \ket{1}$ transitions can be driven much faster than the near-degeneracy of the $\ket{1}$ and $\ket{2}$ states would imply. Because our architecture-agnostic model does not capture this matrix-element selection rule, it artificially penalizes the integer fluxonium in favor of circuits with nondegenerate $\ket{1}$ and $\ket{2}$ states.

\subsection{Three Node Circuits}
There are 80 unique three-node circuits. For each base graph, \cref{tab:optim_bg} lists the four highest-scoring circuits. In \cref{app:more_optim_results}, we also group the highest-scoring circuits by number of junctions, number of flux loops, and number of compact modes.

\begin{table}[ht]
\setlength{\tabcolsep}{3.5pt}
\centering
\begin{tabular}{ccccc}
\toprule
Base Graph & Circuit & Class & $N_{\rm gate}$ (Mean) & $N_{\rm gate}$ (Std) \\ \toprule
\multirow{4}{*}{\includegraphics[]{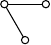}}& \texttt{JL}, \texttt{JCL} & 3-9 & $1.87\cdot10^{5}$ & $5.6\cdot10^{3}$ \\
& \texttt{JCL}, \texttt{JCL} & 3-9 & $1.86\cdot10^{5}$ & $5.0\cdot10^{3}$ \\
& \texttt{JC}, \texttt{CL} & 3-1 & $1.19\cdot10^{5}$ & $2.3\cdot10^{3}$ \\
& \texttt{J}, \texttt{JCL} & 3-13 & $9.53\cdot10^{4}$ & $1.2\cdot10^{3}$ \\
\midrule
\multirow{4}{*}{\includegraphics[]{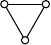}}& \texttt{J}, \texttt{L}, \texttt{CL} & 3-5 & $2.77\cdot10^{5}$ & $1.6\cdot10^{4}$ \\
& \texttt{J}, \texttt{CL}, \texttt{CL} & 3-6 & $2.50\cdot10^{5}$ & $1.1\cdot10^{4}$ \\
& \texttt{J}, \texttt{JL}, \texttt{L} & 3-11 & $2.39\cdot10^{5}$ & $1.8\cdot10^{4}$ \\
& \texttt{JC}, \texttt{CL}, \texttt{CL} & 3-6 & $2.38\cdot10^{5}$ & $1.1\cdot10^{4}$ \\
\bottomrule
\end{tabular}
 \caption{Four highest-scoring three-node circuits for each base graph. The “Base Graph” column shows the circuit topology, and the “Circuit” column lists the elements on each edge, separated by commas. Note that \texttt{J} edges include an intrinsic capacitance.} We also give the corresponding circuit class from \cref{tab:comparison_by_var}.
\label{tab:optim_bg}
\end{table}

Circuits belonging to the $\Delta$ base graph are the highest performing. Their optimized $N_{\rm gate}$ values exceed the two-node benchmark, $N_{\rm gate}\approx1.9\times10^5$, by factors of approximately $1.2$--$1.5$, corresponding to a $20\%$--$50\%$ improvement. However, their larger Monte Carlo variance indicates greater sensitivity to variations in the optimized component parameters.

\begin{figure}[h]
\centering 
\includegraphics[width=0.99\columnwidth]{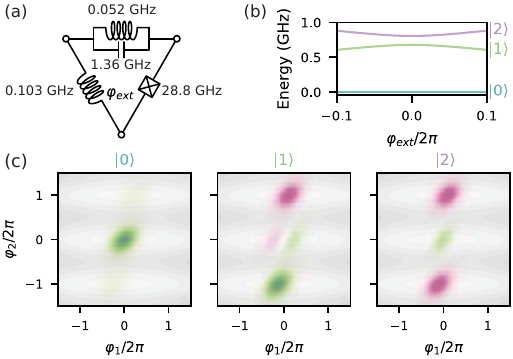}
\caption{
The top-performing circuit ``\texttt{J}, \texttt{L}, \texttt{CL}''. Panel (a) shows the circuit diagrams with parameter values. All junctions also include an intrinsic capacitance $E_{C_J} = 10$ GHz. (b) shows the flux dispersion about the optimized operating point. (c) shows the wavefunctions of the lowest three energy eigenstates, multiplied by a global phase so that the peak of the $\ket{0}$ state is positive and real. Green denotes positive amplitude and pink negative amplitude. The gray background shows contours of the potential energy landscape, with white for low energy and dark gray for high energy. In the optimized ``\texttt{J}, \texttt{L}, \texttt{CL}'' circuit, the wavefunctions appear fluxonium-like in one direction and harmonic-oscillator-like in the other.}
\label{fig:new_wf_1}
\end{figure}

We now examine the highest-performing circuit, \texttt{J}, \texttt{L}, \texttt{CL}, in more detail. It belongs to Hamiltonian class 3-5 and consists of a light extended mode coupled to a heavy harmonic mode.  As shown in \cref{fig:new_wf_1}(a-c), its wavefunctions appear fluxonium-like along the extended coordinate and harmonic-oscillator-like along the other. Its flux dispersion is also fluxonium-like, but with reduced sensitivity near the optimized operating point.

\begin{figure}[h]
\centering 
\includegraphics[width=0.99\columnwidth]{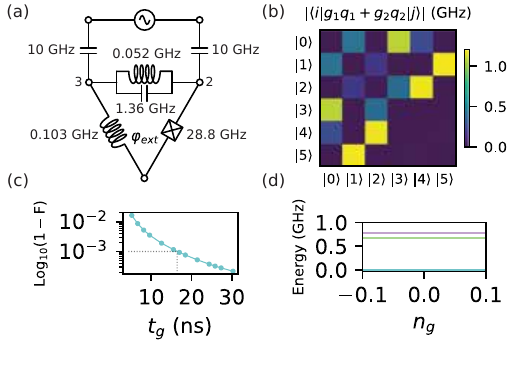}
\caption{(a) Proposed coupling scheme for driving gates on the top-performing circuit ``\texttt{J}, \texttt{L}, \texttt{CL}.'' In the ideal case, the voltage source couples only to the harmonic mode, but capacitance to ground also induces coupling to the extended mode. (b) Magnitudes of the drive-Hamiltonian matrix elements $H_d$ for the lowest five energy levels. Transitions near the qubit frequency (0-2 and 1-3) are forbidden, allowing fast gates with little leakage. (c) Fidelity versus gate time for an optimized monotone pulse with a Gaussian envelope under coherent dynamics. A fidelity of 99.9\% is reached at about 16.4 ns, compared with 18.6 ns predicted by the gate-time model. (d) Dependence of the first and second excited-state energies on the external charge offset. Both energies are insensitive to this parameter.}
\label{fig:new_circuit_drive}
\end{figure}

An explicit coupling scheme for driving single-qubit gates in the top-performing circuit, ``\texttt{J}, \texttt{L}, \texttt{CL},'' is constructed and shown in \cref{fig:new_circuit_drive}(a). 
In principle, the voltage source could be replaced by a resonator for dispersive readout. In the ideal case, with no capacitance to ground, the source couples only to the harmonic mode $q_1$. The coupling capacitors add mass to this already heavy mode and do not meaningfully change the system. When capacitance to ground is included, however, the drive also couples to the extended mode. With dimensionless charge operators $q_i = Q_i/2e$ and charge offset $n_g = C_cV/2e$, the drive Hamiltonian is
\begin{align}
H_{\rm drive} = n_g(g_1q_1 + g_2q_2),
\end{align}
with $g_1 \approx 2.45$ GHz and $g_2 \approx 0.87$ GHz. Here $q_1$ and $q_2$ are the charge operators for the harmonic and extended modes, respectively. See \cref{app:quantization_details} for details of the circuit quantization.

\Cref{fig:new_circuit_drive}(b) shows the direct transition $|0\rangle \leftrightarrow |1\rangle$ has a large matrix element, while nearby transitions in frequency (0-2 and 1-3) are forbidden. This allows fast gates with little leakage. We optimize single-qubit $X$ gates with durations between 5 and 30 ns using a monotone carrier modulated by a Gaussian envelope. In \cref{fig:new_circuit_drive}(c), we plot the infidelity $1-F$ versus gate time $t_g$, where the coherent fidelity $F$ is the average gate fidelity \cite{pedersen_fidelity_2007}. A fidelity of 99.9\% is reached at about 16.4 ns, close to the 18.6 ns predicted by our naive gate-time model. The first and second excited-state energies are insensitive to the new control parameter, see \cref{fig:new_circuit_drive}(d). A key limitation of this design is the requirement for a large inductor and high energy junction with small parallel capacitances.

Our circuit-screening procedure makes several simplifying assumptions. In particular, the original optimization neglected inductor parasitic capacitance, and the Monte Carlo analysis did not vary junction capacitance or capacitance to ground. We therefore added a nominal $0.6$ fF inductor capacitance~\cite{manucharyan_thesis:2012} and reoptimized the four leading three-node circuits in \cref{tab:optim_bg}. We then repeated the fabrication-variation analysis for these circuits. The new analysis includes variations in the inductor, junction, and ground capacitances, together with all previously considered parameter variations; see \cref{tab:re_optim_bg}.

\begin{table}[htb]
\setlength{\tabcolsep}{3.5pt}
\centering
\begin{tabular}{ccccc}
\toprule \vspace{1pt}
Base Graph & Circuit & Class & $N_{\rm gate}$ (Mean) & $N_{\rm gate}$ (Std) \\ \toprule
\multirow{4}{*}{\includegraphics[]{3_node_bg2.pdf}}& \texttt{J}, \texttt{L}, \texttt{CL} & 3-5 & $2.70 \cdot10^{5}$ & $1.5\cdot10^{4}$ \\
& \texttt{J}, \texttt{CL}, \texttt{CL} & 3-6 & $2.51\cdot10^{5}$ & $1.1\cdot10^{4}$ \\
& \texttt{J}, \texttt{JL}, \texttt{L} & 3-11 & $1.90\cdot10^{5}$ & $1.1\cdot10^{4}$ \\
& \texttt{JC}, \texttt{CL}, \texttt{CL} & 3-6 & $2.36\cdot10^{5}$ & $0.9\cdot10^{4}$ \\
\bottomrule
\end{tabular}
 \caption{Reoptimized versions of the top four circuits from \cref{tab:optim_bg}, including inductor parasitic capacitance.  In addition to the parameter variations considered in the original Monte Carlo analysis, we individually vary inductor capacitances, junction capacitances, and ground capacitances with the same Gaussian sampling as before. The inductor, junction, and ground capacitance distributions are centered on $C_L=0.6$ fF, $C_J=1.9$ fF, and $C_g=1$ fF, respectively.}
\label{tab:re_optim_bg}
\end{table} 

Three of the four leading circuits perform comparably. The third-ranked \texttt{J}, \texttt{JL}, \texttt{L} circuit does not contain an explicit capacitor. Unsurprisingly, it is most sensitive to the newly included variations in the inductor, junction, and ground capacitances, which reduce performance by approximately $20\%$. Had they been included in the original optimization, this circuit would likely not have ranked among the top four candidates.
The broader lesson is that relevant effects should be included before optimization. If omitted, they introduce sensitivities and degrade the performance of the optimized design.

\section{Conclusion}\label{sec:conc}

We have enumerated all nonlinear superconducting circuits of up to five nodes built from capacitors, inductors, and Josephson junctions. These circuits could serve as amplifiers~\cite{BlasGrimGirvWallr2021}, current mirrors~\cite{kitaev_protected_2006, weiss_spectrum_2019}, or qubits~\cite{BlasGrimGirvWallr2021}.

As a proof of concept, we used a naive optimization metric to search over all unique three-node circuits to find a good qubit. The best three-node circuit offered an approximate 50\% improvement in the number of gates performed within a coherence time relative to our optimized fluxonium circuit. Because three-node circuits have at most two modes, the range of possible Hamiltonians is limited, and so is the improvement in gate count. A more flexible objective function (e.g., considering a variety of logical states) may reveal better three-node qubits.

Four-node circuits, with richer mode structure and couplings, may outperform three-node circuits, but at the cost of greater complexity. Extending our search for the best qubit to four-node circuits will require more efficient modeling and optimization. \verb|SQcircuit| now includes automatic differentiation, enabling gradient-based methods \cite{rajabzadeh_2024}. Other efforts have added automatic differentiation or GPU support to superconducting-qubit modeling \cite{guilmin2024dynamiqs, XantheComm}, and additional methods, such as tensor networks \cite{di_paolo_efficient_2021}, may also be needed. More advanced numerical methods will be crucial for studying larger circuits. Accurate noise models for larger circuits, such as those proposed by \citet{dravid2026}, would improve predictions as well.

Our enumeration procedure extends readily to other two-terminal devices. For example, we have enumerated all four-node circuits that include a quantum phase-slip junction (QPS) \verb|Q|, built from the components
\begin{align*}
\texttt{C},\, \texttt{J},\, \texttt{L},\, \texttt{Q},\, &\texttt{JC},\, \texttt{CL},\, \texttt{QC},\, \texttt{JL},\,\texttt{JQ},\, \texttt{QL},\, \\ 
\texttt{JCL},\, \texttt{JQC}&, \, \texttt{QCL}, \, \texttt{JQL}, \textrm{and}\, \texttt{JQCL}.
\end{align*}
See \cref{app:subtleties} for the resulting circuit counts. At present, systematic circuit quantization with QPS elements remains an active area of research, and no software exists to carry it out automatically 
~\cite{osborne_symplectic_2023, parra-rodriguez2023}. Thus we cannot construct Hamiltonian classes or search over circuits that include QPS devices. Similarly, extending the enumeration to three- or higher-terminal or non-reciprocal devices (e.g., gyrators) would be much harder because the underlying graph theory is more complex.

Perhaps the most novel contribution of this work is the development of Hamiltonian classes.
Enumeration identifies which unique circuits exist, while Hamiltonian classes organize which circuits realize the same underlying dynamical structure.
Hamiltonian classes extend the notion that a single circuit can behave differently across parameter regimes by showing that different circuits can also behave similarly.
This allowed us to reduce hundreds of millions of circuit configurations down to a few thousand  Hamiltonian classes. 
Moreover, Hamiltonian classes allowed us to identify all circuits up to five nodes whose nonlinear coordinates are necessarily delocalized. Such collective mode structure appears in several protected-qubit designs and provides a natural criterion for identifying promising circuit families.

Our Hamiltonian enumeration results suggest a Mendeleev-like table for multimode circuits. The Mendeleev table, or periodic table, organizes elements by atomic number and reveals trends: elements in the same group often show similar chemical behavior. In superconducting qubits, the family of \texttt{JCL} circuits can be represented as points in the two-dimensional ``table'' defined by $E_L/E_C$ and $E_J/E_C$~\cite{hyyppa_unimon_2022,kalacheva_kinemon_2023}. In this picture, each point represents a different qubit. Such plots provide a simple framework for understanding the differences among most superconducting qubits built today~\cite{DevoretSchoelkopf2013}. The key idea is that varying circuit parameters produces distinct behaviors, and this analogy works well for single-mode circuits.

Our framework suggests a different way to classify multimode circuits. Harmonic, compact (i.e., periodic), and extended modes can be viewed as ``atoms,'' and different circuits as ``molecules'' built from them. Varying circuit parameters then produces different ``species'' or isotopes of those atoms. In this expanded analogy, the atoms (modes), bonds (inter-mode couplings), and species (circuit parameters) together determine the behavior of a superconducting circuit ``molecule.''

\noindent {\em Acknowledgements}: The authors thank Joe Aumentado, Juan Camilo, Sai Pavan Chitta, Luca Corradini, Xanthe Croot, Robert W. Erickson, Peter Groszkowski, Jens Koch, Florent Lecocq, Dragan Maksimovic, Zachary Parrott, Taha Rajabzadeh, Javad Shabani, Thomas Smith, Alex Song, and John Teufel for helpful discussions. 
EW was supported by the National Science Foundation Graduate Research Fellowship under Grant No. 2040434.
MB was supported through CU Boulder's Discovery Learning Apprenticeship (DLA) program.
JC and ZL were supported by the National Science Foundation through a CAREER award ECCS-2240129.

\noindent {\em AI tools statement:} ChatGPT and Claude were used during the later stages of the project to assist with coding and editing. The authors reviewed all AI-generated code and text before use.

\appendix

\section{Subtleties in the Enumeration}\label{app:subtleties}
Enumerating all possible circuits is a formidable task. Although we have carried out this enumeration, it relies on several assumptions. In this appendix, we discuss subtleties that fall outside the scope of the main study and may affect the circuits we enumerate. These include stray capacitances and inductances, readout resonators, transmission lines, and Purcell filters. The list is not exhaustive, and no single study could address all such complications.

\subsection{Ground Placement and Circuit Equivalence}\label{sec:gnd}
Our enumeration procedure does not assign any nodes to ground, although the choice of ground can significantly change a circuit's Hamiltonian. When modeling circuits in \cref{sec:optimization}, we choose to include capacitive coupling between each node and ground.

\begin{figure}[ht]
\centering 
\includegraphics[width=0.99\columnwidth]{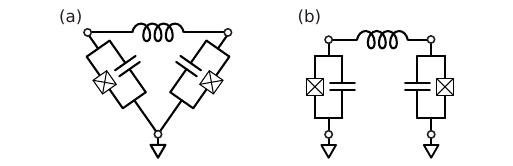}
\caption{Illustration of how the number and placement of ground nodes can induce circuit isomorphism. Circuit (a) has three nodes, while circuit (b) has four. The two circuits become equivalent when the indicated one node in (a) and two nodes in (b) are grounded.}\label{fig:ground_transformation}
\end{figure}

Ground placement can also change the effective circuit topology. Consider the two circuits in \cref{fig:ground_transformation}(a) and (b), which have three and four nodes, respectively. If we ground one node in \cref{fig:ground_transformation}(a) and two nodes in \cref{fig:ground_transformation}(b), both reduce to a circuit that resembles galvanically coupled transmons. As a result, including grounds can cause our procedure to over-count unique circuits. For a graph with $N$ nodes, there are
$\sum_{k=1}^{N-1}\binom{N}{k} = 2^N - 2$
ways to ground one or more nodes. One could enumerate these possibilities and identify equivalent configurations by adding a ground ``color'' to the graph-isomorphism procedure described in \cref{sec:CirEnumAlg}. However, grounding multiple nodes effectively reduces the number of nodes in the circuit. These reduced circuits are therefore not new configurations and already appear in the enumeration of smaller circuits.

\subsection{Ambiguities Related to SQUIDs and Linearization}\label{sec:SQUIDs}
\begin{figure}[ht]
\centering 
\includegraphics[width=0.99\columnwidth]{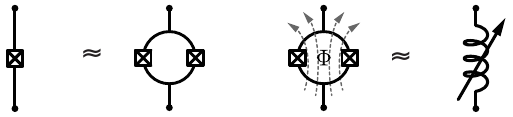}
\caption{Flux-tunable elements not included explicitly in our enumeration.}\label{fig:SQUID_amb}
\end{figure}

We did not include superconducting quantum interference devices (SQUIDs) or tunable inductors built from flux-biased SQUIDs in our enumeration. In a sense, any Josephson junction in our enumeration could be viewed as a split junction, i.e., a SQUID. Likewise, any linear inductor could be viewed as a linearized SQUID; see \cref{fig:SQUID_amb}. These distinctions matter at the Hamiltonian level because SQUID-based elements introduce additional tunable terms, such as $(\phi-\phi_{\rm ext})^2$ and $\cos(\phi-\phi_{\rm ext})$.

Flux bias can also be used to engineer effective cubic or quartic terms \cite{Frattini_snail_2017, Quarton2021}. A standard example is the SNAIL \cite{Frattini_snail_2017}. The SNAIL consists of several Josephson junctions in a series-parallel arrangement. By tuning the external flux, one can isolate selected powers of the phase operator, $\phi^n$. To include such effects, one could introduce new effective dipole elements into the enumeration. This is a direction that could reveal new circuit classes, but it lies beyond the scope of our present procedure.

\subsection{Charge and Flux Bias} \label{sec:charge_flux_bias}
\begin{figure}[ht]
\centering 
\includegraphics[width=0.99\columnwidth]{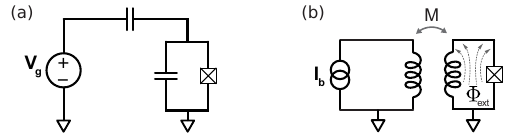}
\caption{Examples of (a) a charge-bias circuit and (b) a flux-bias circuit, neither of which is included explicitly in the enumeration.}\label{fig:flux_n_charge_bias}
\end{figure}

We do not include charge and flux biases in the enumeration itself, although they do appear in the optimization over unique circuits; see \cref{fig:flux_n_charge_bias}. These biases can distinguish circuits that otherwise belong to the same Hamiltonian class, as defined in \cref{sec:HamEnum}.

To see this, consider the circuit Hamiltonian
\begin{align}
    H = \frac{1}{2}(Q - Q_g)^T C^{-1}(Q - Q_g) + U(\Phi, \Phi_{\rm ext}),
\end{align}
where $Q$ is the vector of node charges, $\Phi$ is the vector of node fluxes, $C^{-1}$ is the inverse capacitance matrix, and $Q_g$ is a vector of gate charges. We can identify a separate circuit and coupling Hamiltonian:
\begin{align}
    H_{\rm circuit} &= \frac{1}{2}Q^T C^{-1}Q + U(\Phi) \\
    H_{\rm coupling} &= Q^T C^{-1} C_V V + U(\Phi, \Phi_{\rm ext}).
\end{align}
In \cref{sec:HamEnum}, we classify circuits using only $H_{\rm circuit}$. This is equivalent to setting $Q_g = 0$ and  $\Phi_{\rm ext} = 0$ in the full Hamiltonian. Circuits in the same Hamiltonian class may still differ in $H_{\rm coupling}$. Those differences determine how the system can be controlled and may also affect decoherence. 
A classification that also incorporates these bias-dependent coupling Hamiltonians is left to future work.

\subsection{Circuit Elements Not Considered} \label{subsec:elems considered}
\begin{figure}[ht]
\centering 
\includegraphics[width=0.99\columnwidth]{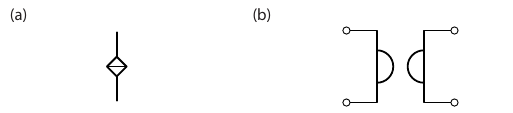}
\caption{QPS elements and multiport/multiterminal devices.}\label{fig:QPS_mulitport}
\end{figure}

We did not include several elements that appear in the superconducting-circuit literature, such as resistors and three-terminal devices like circulators. Two cases deserve further comment: quantum phase-slip junctions and multiport nonreciprocal devices; see \cref{fig:QPS_mulitport}.

A quantum phase-slip junction (QPS)~\cite{Mooij_QPS_2006} is effectively a nonlinear capacitor and is the electrical dual of the Josephson junction. Although QPS has not yet been realized unequivocally in experiment, we excluded it from the main study for a more practical reason: modern superconducting-circuit analysis packages do not support it. That would sharply limit the analysis of the resulting circuits. Even so, adding a new two-terminal device is straightforward at the level of \textit{circuit} enumeration. To begin exploring circuits with QPS, we carried out a secondary enumeration of circuits with up to four nodes using all fifteen combinations of capacitor (\verb|C|), inductor (\verb|L|), Josephson junction (\verb|J|), and QPS (\verb|Q|).
\begin{align*}
\texttt{C},\, \texttt{J},\, \texttt{L},\, \texttt{Q},\, &\texttt{JC},\, \texttt{CL},\, \texttt{QC},\, \texttt{JL},\,\texttt{JQ},\, \texttt{QL},\, \\ 
\texttt{JCL},\, \texttt{JQC}&, \, \texttt{QCL}, \, \texttt{JQL}, \textrm{and}\, \texttt{JQCL}.
\end{align*}
\Cref{tab:qps_enum} gives the resulting number of unique circuits.

\begin{table}[ht]
\setlength{\tabcolsep}{15pt}
\begin{tabular}{@{}ccc@{}}
\toprule
Nodes \,\,      &  Circuits&  Unique Nonlinear Circuits  \\ \toprule
2                  & 15          &   12          \\ \midrule
3                  & 3600        &   736         \\ \midrule
4                  & 12,258,000    &   712,682 \\ \bottomrule
\end{tabular}
 \caption{Number of enumerated circuits as a function of node count for circuits built from \texttt{C}, \texttt{L}, \texttt{J}, and \texttt{Q}. Column 1 gives the number of nodes. Column 2 gives the naive count, $15^{N_E}$, summed over all base graphs for that node count (see \cref{fig:basegraph_grid}). Column 3 gives the number of unique nonlinear circuits containing at least one Josephson junction or one QPS element.}\label{tab:qps_enum}
\end{table}

Nonreciprocal devices such as circulators and gyrators raise two separate issues: nonreciprocity and multiport structure. The first is that consistent quantization of nonreciprocal devices has only recently been developed~\cite{parra-rodriguez2023}, and current software packages do not yet support it. The second is that multiport devices would add substantial overhead to the enumeration itself. Their circuit topologies would need to be represented by directed hypergraphs rather than undirected graphs. That would increase the number of possible configurations and prevent us from using widely available precomputed lists of nonisomorphic simple graphs, since comparable lists for hypergraphs are not generally available.

\subsection{Eliminating Single-Node Capacitive and Inductive Islands}\label{app:details:islands}

We do not consider circuits that contain a single node connected only to capacitors or only to inductors. In such cases, one can construct an equivalent circuit with one fewer node and the same equations of motion. We show below how to identify the required effective capacitance or inductance from the circuit Lagrangian. This construction is equivalent to a generalized delta-Y, or star-mesh, transformation \cite{rosen_new_1924}.

Consider a subcircuit of $n$ capacitors, $[C_1, C_2, \dots, C_n]$, that meet at a central node. Additional capacitive connections may already exist between the outer terminals.

The central-node variable is a free variable, but it is coupled to the outer nodes, which may themselves be dynamical. We now follow the procedure in \cref{app:var_trans:decouple} to decouple this free variable and derive reduced dynamics that do not depend on the central node.

We first split the node flux variables into two parts,
\begin{align*}
\phi = \begin{pmatrix}
    \phi_1 \\
    \phi_2
\end{pmatrix}.
\end{align*}
where $\phi_1$ contains the outer nodes and $\phi_2$ contains the inner node variable. We likewise partition the capacitance matrix into blocks for the outer nodes $C_{11}$, inner nodes $C_{22}$, and their couplings $C_{21}, C_{12}$:
\begin{align*}
    C &= \begin{pmatrix}
    C_{11} & C_{12} \\
    C_{21} & C_{22}
\end{pmatrix}.
\end{align*}
Given the structure of the circuit,
\begin{align}
\begin{split}
    C_{11} &= \rm diag([C_1, C_2, \dots, C_n]) + C_0 \\
    C_{22} &= \sum_{i=1}^n C_i = C_\Sigma \\
    C_{21} &= C_{12}^T = \rm row([C_1, C_2, \dots, C_n]),
\end{split}
\end{align}
where $\rm diag$ denotes a diagonal matrix and $\rm row$ a row vector. Here $C_i$ is the capacitance between outer node $i$ and the central node, while $C_0$ contains the capacitances that connect the outer ports to one another or to the rest of the circuit.

The modified capacitance matrix
\begin{align}
    \tilde{C} = C_{11} - C_{12}C_{22}^{-1}C_{21} = C/C_{22}
\end{align}
defines a reduced system dynamics. To calculate $\tilde{C}$, first consider the entries
\begin{align*}
    [C/C_{22}]_{ij} = [\begin{pmatrix}
        C_1 \\
        \vdots \\
        C_n
    \end{pmatrix} \frac{1}{C_\Sigma}\begin{pmatrix}
        C_1 & \dots & C_n
    \end{pmatrix}]_{ij} = \frac{C_iC_j}{C_\Sigma}.
\end{align*}
The diagonal entries are therefore
\begin{align}\label{eq:reduced_diagonal}
    [\tilde{C}]_{ii} = [C_0]_{ii} + C_i - \frac{C_i^2}{C_\Sigma} = [C_0]_{ii} + \frac{\sum_{i\neq j} C_iC_j}{C_\Sigma}
\end{align}
and off-diagonal entries are
\begin{align}
    [\tilde{C}]_{ij} = [C_0]_{ij} - \frac{C_iC_j}{C_\Sigma}.
\end{align}

It follows that the same Lagrangian can be realized by a circuit with no inner node and an added capacitance $\frac{C_iC_j}{C_\Sigma}$ between outer nodes $i$ and $j$. In the special case of one inner node and one outer node, that is, a dangling capacitor or inductor attached to a single node, \cref{eq:reduced_diagonal} reduces to
$[\tilde{C}]_{ii} = [C_0]_{ii} + C_i - C_i^2/C_\Sigma = [C_0]_{ii}$. The dangling element therefore does not affect the dynamics, as no current can flow through that branch.

We omit the corresponding derivation for a node connected only to inductors, as it is identical to the capacitive case.

\section{Variable Transformation Details}\label{app:var_trans}
\subsection{Choice of ``\Canonical'' Coordinates}
The goal of this section is to define a set of ``\canonical'' coordinates for a general superconducting circuit made of capacitors, inductors, and Josephson junctions. Our method generates a transformation to bring any circuit Lagrangian into a standard form, which we use to categorize circuits into ``Hamiltonian classes.'' 
Hamiltonian classes consist of Hamiltonians with the same functional form but different parameter values. 

The definition of our \canonical\ coordinates is motivated by a desire for minimally coupled circuit modes, without a dependence on underlying circuit parameters. The underlying transformation algorithm is flexible and could be used to generate different choices of ``\canonical'' coordinates.

For a general circuit, we impose the following requirements.
\begin{enumerate}
\item[(1)] Maximize the number of non-extended modes, identifying compact, harmonic, and non-dynamical modes wherever possible.
\item[(2)] Non-dynamical modes must not couple to dynamical modes.
\item[(3)] All dynamical coordinates must be parameter-independent linear combinations of node variables.
\item[(4)] Compact coordinates form an $n$-dimensional cubic unit cell, with charges quantized in units of Cooper pairs. Thus, the lattice vectors of the periodic potential are the standard basis vectors \cite{Rajabzadeh2023analysisofarbitrary}.
\end{enumerate}
Within the constraints of the above requirements, some freedom in the choice of coordinates remains. We fix a unique choice by imposing the following criteria, in order of priority:

\begin{enumerate}
\item[(a)] Minimize nonlinear inter-mode coupling, measured by the off-diagonal support of 
$J = \sum_i E_{J,i}\,\omega_{J,i}\omega_{J,i}^T$, where $\omega_{J,i}$ denotes the $i$th junction incidence vector.
\item[(b)] Minimize linear inter-mode coupling, defined by the number of nonzero off-diagonal entries in the transformed inverse capacitance ($C^{-1}$) and susceptance ($L^*$) matrices. In case of a tie, prioritize minimizing inductive coupling. As a further tiebreaker, prefer transformations with fewer off-diagonal terms when all circuit parameters are taken to be identical.
\item[(c)] Ensure integer arguments in cosine potentials of extended modes (e.g., $\cos(2\varphi)$, not $\cos(1.45\varphi)$). Coefficients are chosen to minimize the $\ell_1$ norm of the transformed junction incidence matrix.
\end{enumerate}

Finally, we impose a preferred ordering of variable labels. Specifically, we order the coordinate vector as
$\Phi = (\Phi^{c}, \Phi^{e}, \Phi^{h}, \Phi^{\rm fr,fz,cy})^T$,
with a vector of compact variables first, followed by extended, harmonic, and non-dynamical variables.

\subsection{Generating a ``\Canonical'' Transformation}

The requirements of our ``\canonical'' coordinates motivate a sequence of three linear transformations. The first transformation splits modes into different types and addresses requirements (1) and (4). The second transformation enforces requirement (2) by removing any residual coupling between dynamical and non-dynamical variables. 
Both steps can be carried out without violating the parameter-independence requirement of (3).
Finally, the third transformation performs the minimization described in (a) and (b), again subject to the parameter dependence requirement of (3). 
An implementation of this transformation is included in our code \cite{sircuitenum_repo}.

\subsubsection{Separate Mode Types}

For a circuit with capacitance matrix $C$, susceptance matrix $L^*$, and junction incidence matrix $\omega_J$, the Lagrangian is written in terms of node flux variables $\Phi$ as
\begin{align}\label{eq:node_lagrangian}
    \mathcal{L}(\Phi, \dot \Phi) = \frac{1}{2}\dot\Phi^TC\dot\Phi &- \frac{1}{2}\Phi^TL^*\Phi  \nonumber \\ &+ \sum_iE_{J,i}\cos\left(\frac{2\pi}{\Phi_0}\omega_{J,i}^T\Phi\right),
\end{align}
neglecting charge offset and external flux. Here $\Phi_0$ is the superconducting magnetic flux quantum. $E_{J,i}$ and $\omega_{J,i}$ are the Josephson energy and incidence matrix column corresponding to junction $i$. Each $\omega_J$ column represents a Josephson junction, with a $\pm1$ at entries corresponding to connected nodes. For example, a three-node circuit with a junction connecting nodes two and three would have $\omega_J = \{0,1,-1\}$ or $\omega_J = \{0,-1,1\}$. Because cosine is an even function, the choice of head/tail does not matter, though consistency is key when considering time-dependent flux drives \cite{Rajabzadeh2023analysisofarbitrary}. Note that the linear terms can also be expressed with incidence matrices, such as $\dot\Phi^TC\dot\Phi = \sum_i(C_i\omega_{C,i}^T\dot \Phi)^2$.

We define a new set of coordinates $\tilde{\Phi}$ by a linear transformation of the node flux variables,
$\tilde{\Phi} = Z^{-1}\Phi$.
Although the inverse may appear unconventional, this choice ensures that the capacitance matrix $C$, the susceptance matrix $L^*$, and the junction incidence vectors $\omega_J$ transform with $Z$. This is seen by substituting $\Phi = Z\tilde{\Phi}$ into \cref{eq:node_lagrangian}, which yields
\begin{align}\label{eq:tilde_lagrangian}
    \mathcal{L}(\tilde \Phi, \dot{\tilde\Phi}) &= \frac{1}{2}\dot{\tilde\Phi}^TZ^TCZ\dot{\tilde\Phi} - \frac{1}{2}\tilde \Phi^TZ^T L^*Z\tilde \Phi  \nonumber \\ & \quad + \sum_iE_{J,i}\cos\left(\frac{2\pi}{\Phi_0}\omega_{J,i}^T Z \tilde \Phi\right)
\end{align}
and can now be written in terms of transformed matrices
\begin{align}
    &\tilde C = Z^TCZ, \quad \tilde{L}^* = Z^TL^*Z, ~~\textrm{and}~~ \tilde \omega_J^T = \omega_J^T Z ~.
\end{align}

For circuits with capacitors, inductors, and Josephson junctions with intrinsic capacitance, six distinct modes arise\footnote{Consistently treating Josephson junctions without intrinsic capacitance can lead to multi-valued constraints and remains an area of active research~\cite{rymarz_consistent_2023, Eguisquiz_Reply_Consistent_2025, Eguisquiz_Comment_Consistent_2025}.}. Three are dynamical: compact (cosine potential), extended (cosine + quadratic potential), and harmonic (quadratic potential). Three are non-dynamical: free (kinetic only), frozen (potential only), and cyclic (absent from the Lagrangian).

We define coordinates $\Phi_1$ = $\{\Phi_1^{\rm c}, \Phi_1^{\rm e}, \Phi_1^{\rm h},        \Phi_1^{\rm fr}, \Phi_1^{\rm fz} , \Phi_1^{\rm cy}\}^T$ that contain blocks corresponding to different mode types.\footnote{A circuit has as many modes as active circuit nodes, so for any given circuit, many of these blocks will be of dimension zero.} The corresponding transformation can be written as $\Phi = Z_1\Phi_1$, or
\begin{align}
     &\begin{pmatrix}
        \phi_0 \\ \phi_1 \\ \vdots \\ \phi_{n-1} \\ \phi_n
    \end{pmatrix}
    = Z_1 
    \begin{pmatrix}
       \Phi_1^{\rm c} \\\Phi_1^{\rm e} \\ \Phi_1^{\rm h} \\
       \Phi_1^{\rm fr} \\\Phi_1^{\rm fz} \\ \Phi_1^{\rm cy}
    \end{pmatrix}.
\end{align}
Here $\phi_i$ denotes a circuit node flux variable. The columns of the transformation $Z_1$ impose restrictions on the nature of the corresponding variables. To see this, consider $Z_1$ as blocks corresponding to different variable types
\begin{align}
Z_1 
    \begin{pmatrix}
       \Phi_1^{\rm c} \\\Phi_1^{\rm e} \\ \Phi_1^{\rm h} \\
       \Phi_1^{\rm fr} \\\Phi_1^{\rm fz} \\ \Phi_1^{\rm cy}
    \end{pmatrix}&= \begin{pmatrix}
        Z_1^{\rm c} & Z_1^{\rm e} & Z_1^{\rm h} & Z_1^{\rm fr} & Z_1^{\rm fz} & Z_1^{\rm cy}
    \end{pmatrix}
    \begin{pmatrix}
       \Phi_1^{\rm c} \\\Phi_1^{\rm e} \\ \Phi_1^{\rm h} \\
       \Phi_1^{\rm fr} \\\Phi_1^{\rm fz} \\ \Phi_1^{\rm cy}
    \end{pmatrix} \nonumber \\
    &= (
       Z_1^{\rm c}\Phi_1^{\rm c} +Z_1^{\rm e}\Phi_1^{\rm e} + Z_1^{\rm h}\Phi_1^{\rm h} + \nonumber \\
    & \quad ~~~   Z_1^{\rm fr}\Phi_1^{\rm fr} + Z_1^{\rm fz}\Phi_1^{\rm fz} + Z_1^{\rm cy}\Phi_1^{\rm cy}
    ).
\end{align}
For example, consider the requirements to define a harmonic coordinate. It must not have a cosine potential, which requires $\Phi_1^{\rm h}$ never appear in any $\cos$ term of \cref{eq:node_lagrangian}. Equivalently, $\omega_{J}^TZ\Phi_1 = \omega_{J}^T(
       Z_1^{\rm c}\Phi_1^{\rm c} +Z_1^{\rm e}\Phi_1^{\rm e} + Z_1^{\rm h}\Phi_1^{\rm h} + Z_1^{\rm fr}\Phi_1^{\rm fr} + Z_1^{\rm fz}\Phi_1^{\rm fz} + Z_1^{\rm cy}\Phi_1^{\rm cy}
    )$
must not depend on $\Phi_1^{\rm h}$. This condition is satisfied if and only if
\begin{align}\label{eq:omegaj_null}
    \omega_J^TZ_1^{\rm h} = 0 \quad \textrm{or} \quad \textrm{col}(Z_1^{\rm h}) \in \textrm{null}(\omega_J^T),
\end{align}
where the notation $\textrm{col}(A)$ denotes the set of all linear combinations of the columns of $A$, and $\textrm{null}(A)$ means the set of vectors $x$ for which $Ax=0$.
So \cref{eq:omegaj_null} means all columns of $Z_1^{\rm h}$ are inside the nullspace of $\omega_J^T$. A harmonic mode must also have a quadratic potential and momentum, which requires
\begin{align}
    \textrm{col}(Z_1^{\rm h}) \notin \textrm{null}(\omega_L^T) \quad \textrm{and} \quad \textrm{col}(Z_1^{\rm h}) \notin \textrm{null}(\omega_C^T)~,
\end{align}
for inductor and capacitor incidence matrices $\omega_L$ and $\omega_C$.

\begin{figure}[t!]
\centering
\resizebox{0.85\columnwidth}{!}{\begin{tikzpicture}

\draw[thick] (0,0) circle (4.0);
\node at (3.3, 3.3) {\large \(\mathbb{R}^{n_{\text{nodes}}}\)};
\node at (0,3.4) {\footnotesize Extended};
\node at (0, 3.4-0.4) {\footnotesize(general)};

\draw[thick, blue] (1.5*-0.866,1.5*0.5) circle (2.1);
\node[blue] at (-1.45, 2.4) {\normalsize \({\mathrm{null}(\omega_J^T)}\)};
\node at (2*-0.866,2*0.5) {\footnotesize Harmonic};

\draw[thick, red] (1.5*0.866,1.5*0.5) circle (2.1);
\node[red] at (1.45, 2.4) {\normalsize \({\mathrm{null}(\omega_L^T)}\)};
\node at (2*0.866,2*0.5) {\footnotesize Compact};

\draw[thick, green!70!black] (1.5*0,1.5*-1) circle (2.1);
\node[green!70!black] at (0,-3.25) {\normalsize \(\mathrm{null}(\omega_C^T)\)};
\node at (0, -2) {\footnotesize Extended};
\node at (0, -2.4) {\footnotesize (light)};

\node at (0.0, 1.0) {\footnotesize Free};
\node at (0, 0.0) {\footnotesize Cyclic};
\node at (-1.1, -0.60) {\footnotesize Frozen};
\node at (1.1, -0.60) {\footnotesize Compact};
\node at (1.1, -1.00) {\footnotesize (light)};

\end{tikzpicture}}
\caption{Types of modes in superconducting circuits made of pure capacitors, pure inductors, and Josephson junctions with intrinsic capacitance. The columns of $Z_1^{m}$ for a given mode type $m$ are placed within the null spaces of the incidence matrices $\omega_J^T$, $\omega_L^T$, and $\omega_C^T$. Light indicates a small capacitance or ``mass'' from only the intrinsic junction capacitance.}\label{fig:mode_diagram}
\end{figure}

Our algorithm incorporates the analogous requirements on $\textrm{null}(\omega_J)$, $\textrm{null}(\omega_L)$, and $\textrm{null}(\omega_C)$ shown in \cref{fig:mode_diagram}, which enable the automatic construction of $Z_1$. We describe the procedure below.
\begin{enumerate}
    \item[(1)] Construct incidence matrices $\omega_J, \omega_C, $ and $\omega_L$.
    \item[(2)] Determine the dimension of each zone in \cref{fig:mode_diagram} to identify the modes present.
    \item[(3)] For non-compact modes $m$, select any valid choice of $Z_1^m$. These columns will be redefined later to minimize inter-mode coupling.
    \item[(4)] For compact modes $c$, the desired lattice vectors are achieved when a subset of rows in $\omega_J^TZ_1^c$ is equal to the identity matrix. Beginning with any choice of $Z_1^c$ that respects \cref{fig:mode_diagram}, all $Z_1^c$ with desired lattice vectors can be obtained by inverting different linearly-independent sets of rows of $\omega_J^TZ_1^c$.
\end{enumerate}

\subsubsection{Decouple Non-Dynamical Modes} \label{app:var_trans:decouple}
After the first transformation, dynamical modes may be linearly coupled to non-dynamical modes. It is always possible to define a transformation that decouples them \cite{ding_free-mode_2021}. We can define this secondary transformation $\Phi_1 = Z_2\Phi_2$ in terms of the dynamical (c, e, h) and non-dynamical (fr,fz,cy) blocks of $\Phi_1$
\begin{align}
    \begin{pmatrix}
        \Phi_1^{\rm c,e,h} \\ \Phi_1^{\rm fr,fz,cy}
    \end{pmatrix} = Z_2 \begin{pmatrix}
        \Phi_2^{\rm c,e,h} \\ \Phi_2^{\rm fr,fz,cy}
    \end{pmatrix}.
\end{align}
The only choice of $Z_2$ that does not alter the definitions of dynamical variables is 
\begin{align}
    Z_2 = \begin{pmatrix}
    I & 0 \\
    M & I
\end{pmatrix} ~,
\end{align}
with $M= -C_{22}^{-1}C_{21}$ or $M = -(L^*_{22})^{-1}L^{*}_{21}$ depending on whether the dynamical modes are coupled by charge or flux. Block 1 denotes dynamical modes (c, e, h), and block 2 denotes non-dynamical modes (fr, fz, cy). If coupling occurs through both, the two transformations are applied in series. It can be shown that the decoupling transformation $Z_2$ is unique up to redefining the non-dynamical modes. 

It is worth noting that while the inverse transformation 
\begin{align}
    \begin{pmatrix}
        \Phi_2^{\rm c,e,h} \\ \Phi_2^{\rm fr,fz,cy}
    \end{pmatrix} = \begin{pmatrix}
    \phantom{-}I & 0 \\
    -M & I
\end{pmatrix} \begin{pmatrix}
        \Phi_1^{\rm c,e,h} \\ \Phi_1^{\rm fr,fz,cy}
    \end{pmatrix}
\end{align}
does leave the definition of $\Phi_1^{\rm c,e,h}$ unchanged, it can introduce parameter dependence into the definitions of non-dynamical coordinates.

\subsubsection{Reduce Intermode Coupling}

The final transformation $Z_3$ accounts for all transformations that preserve the mode structure while leaving compact variables unchanged. A new $Z_3$ is generated for each choice of compact variable considered in $Z_1$. The transformation can be written as $\Phi_2 = Z_3\Phi_3$, or
\begin{align}\label{eq:Z3}
    \begin{pmatrix}
        \Phi_2^{\rm c} \\\Phi_2^{\rm e} \\ \Phi_2^{\rm h} \\
       \Phi_2^{\rm fr,fz,cy}
    \end{pmatrix} = \begin{pmatrix}
        I & Z_3^{\rm ce} & 0 & 0 \\
        0 & Z_3^{\rm e} & 0 & 0 \\
        0 & Z_3^{\rm{he}} & Z_3^{\rm h} & 0 \\
        0 & 0 & 0 & I  \\
    \end{pmatrix} 
    \begin{pmatrix}
    \Phi_3^{\rm c} \\\Phi_3^{\rm e} \\\Phi_3^{\rm h} \\
    \Phi_3^{\rm fr,fz,cy}
    \end{pmatrix}.
\end{align}
Note that $Z_3$ acts only on dynamical modes, ensuring they remain decoupled from non-dynamical modes. The blocks $Z_3^{\rm ce}, Z_3^{\rm e}, Z_3^{\rm he},$ and $Z_3^{\rm h}$ are selected to minimize nonzero inter-mode coupling terms while maintaining parameter independence. Using the computer algebra systems \texttt{Sage} ~\cite{sagemath} and \texttt{Singular} ~\cite{Singular_2024} and the SMT solver \texttt{Z3} ~\cite{DeMouraBjornerZ3_2008}, we are able to obtain a provably optimal solution over rational numbers \cite{sircuitenum_repo, CircuitQuantPaper2026}.

A final ambiguity arises because multiple choices of $Z_3$ can produce the same number of coupling terms. For example, consider two inductive couplings among three modes
\begin{align}
    &L_3^{(1)} = \begin{pmatrix}
        1/L_{a} & -1/L_{a}L_{b} & 0 \\
        -1/L_{a}L_{b} & 1/L_{b} & 0 \\
        0 & 0 & 1/L_{c}
    \end{pmatrix}\, \quad \textrm{and} \\
    &L_3^{(2)} = \begin{pmatrix}
        1/L_{a} & 0 & 0 \\
        0 & 1/L_{b} & -1/L_{b}L_{c} \\
        0 & -1/L_{b}L_{c} & 1/L_{c}
    \end{pmatrix}.
\end{align}
Which one should be preferred? Both are equivalent, differing only by an interchange of the first and third modes. We make the arbitrary choice to favor $L_3^{(2)}$.

We choose the column ordering that minimizes a binary string representing the nonzero off-diagonal entries of a square matrix. 
For example, $L_3^{(1)} \rightarrow 100$, while $L_3^{(2)} \rightarrow 001$. Columns must respect the ordering of \cref{eq:Z3} (compact, extended, harmonic, free, frozen, cyclic), so swaps are only allowed within like modes.

After the third transformation, non-dynamical modes are discarded. The circuit Hamiltonian then follows from a Legendre transform, and its structure is used to group circuits with similar mode patterns.

\subsection{Limitations}

This method relies on the standard node flux formalism \cite{Nigg2012,Solgun2014,YouKoch2019,Rasmussen2021} to obtain a circuit Hamiltonian. Node flux methods are well defined for circuits that contain reciprocal elements and a single type of nonlinear element (e.g., they cannot contain both a nonlinear capacitor and nonlinear inductor), as shown in Sec. IV.H of  \cite{osborne_symplectic_2023}. If a quantum phase slip junction (nonlinear capacitor) or non-reciprocal element were included alongside the Josephson junction, more modern geometric circuit quantization methods would be required to systematically obtain circuit Hamiltonians \cite{osborne_symplectic_2023,parra-rodriguez2023}.

We additionally do not consider any emergent dynamics from considering circuit singularities derived from Josephson junctions without intrinsic capacitance. A circuit is considered ``singular'' when its capacitance matrix is not invertible (e.g., when it has a nontrivial null vector). In reality there is always some capacitance between every pair of nodes in a physical circuit, although it will be many orders of magnitude smaller than a dedicated capacitor. Considering these small capacitances as zero allows us to construct a reduced model with fewer variables. Each zero-capacitance island introduces a constraint upon the circuit variables, effectively reducing the degrees of freedom present. Each island also corresponds to a nontrivial null vector in the capacitance matrix. In the case of linear inductors without a parallel capacitance, this constraint is resolved straightforwardly by removing the ``frozen'' variable. However, if the island is shunted by a pure nonlinear inductor, this constraint can be multi-valued, and how to accurately resolve the singularity remains an open question \cite{rymarz_consistent_2023, Eguisquiz_Reply_Consistent_2025}. To avoid this inconsistency, we assume that Josephson junctions have an intrinsic capacitance, which we treat as a small ever-present capacitor in parallel with their nonlinear inductance.

\section{Pseudocode for Constructing a Nonisomorphic Set of Colored Graphs}\label{app:psuedo}
To construct a nonisomorphic set of colored graphs from a collection that may contain isomorphic graphs, we build the set one element at a time, following Algorithm 3 of \cite{Herber:2017aa}. For each candidate graph, we test isomorphism against the graphs already in the ``unique'' set. If it is not isomorphic to any of them, we add it. We perform these isomorphism checks with the Python package \texttt{networkx}. Details of the implementation are available in our GitHub repository \cite{sircuitenum_repo}.

For clarity, we also include pseudocode for this procedure.

\begin{algorithm}[H]
\caption{Constructing a Nonisomorphic Set of Colored Graphs}
\label{alg:nonisomorphic}
\begin{algorithmic}
\Require A set of colored graphs $G = \{G_1, G_2, \dots, G_n\}$
\Require A function \texttt{isomorphic}$(G_1, G_2)$ that tests graph isomorphism
\Ensure A subset $S \subseteq G$ containing one representative from each isomorphism class
\State $S \gets \{\}$
\For{each graph $G_i \in G$}
    \State $u \gets \text{true}$
    \For{each graph $S_j \in S$}
        \If{\texttt{isomorphic}$(G_i, S_j)$}
            \State $u \gets \text{false}$
            \State break
        \EndIf
    \EndFor
    \If{u}
        \State $S \gets S\cup \{G_i\}$
    \EndIf
\EndFor
\State \Return $S$
\end{algorithmic}
\end{algorithm}

\section{Additional Enumeration Results by Number of Junctions, Loops, and Compact Variables}\label{app:additional_results}
In this appendix, we provide three additional tables that summarize the numbers of circuits and Hamiltonian classes by number of Josephson junctions (\cref{tab:granular_overview}),  number of compact variables (\cref{tab:stats_periodic}), and number of flux loops (\cref{tab:granular_overview2}).

For example, there are 2306 unique four-node circuits with two Josephson junctions. This set includes the \zp\ qubit \cite{Gyenis2021} and the Difluxmon \cite{garcia-azorin_2024}. Both arise from the fully connected four-node base graph shown in \cref{fig:basegraph_grid}, which can also be drawn as a triangle with an interior node.

As the number of nodes increases, compact modes become less common among unique circuits. Indeed, 50\%, 45\%, 38\%, and 26\% of unique circuits with 2, 3, 4, and 5 nodes, respectively, contain at least one compact mode. By contrast, the fraction of Hamiltonian classes containing at least one compact variable increases with node count.

\begin{table}[h]
\centering
\setlength{\tabcolsep}{14pt}
\begin{tabular}{cccc}
\toprule
Nodes & JJs & Unique Circuits & Classes \\ \toprule
\multirow{1}{*}{2}& 1 & 4 & 2 \\
\midrule
\multirow{3}{*}{3}& 1 & 20 & 6 \\
& 2 & 40 & 9 \\
& 3 & 20 & 4 \\
\midrule
\multirow{6}{*}{4}& 1 & 744 & 46 \\
& 2 & 2306 & 61 \\
& 3 & 3588 & 59 \\
& 4 & 2852 & 41 \\
& 5 & 1216 & 15 \\
& 6 & 276 & 10 \\
\midrule
\multirow{10}{*}{5}& 1 & 52012 & 614 \\
& 2 & 270895 & 846 \\
& 3 & 819532 & 838 \\
& 4 & 1584528 & 826 \\
& 5 & 1952812 & 616 \\
& 6 & 1745973 & 508 \\
& 7 & 1044878 & 374 \\
& 8 & 401711 & 170 \\
& 9 & 93920 & 73 \\
& 10 & 10688 & 28 \\
\bottomrule
\end{tabular}
\caption{Number of unique circuits and Hamiltonian classes, grouped by node count and number of Josephson junctions.}
    \label{tab:granular_overview}
\end{table}

\begin{table}[h]
\setlength{\tabcolsep}{11pt}
\centering
\begin{tabular}{cccc}
\toprule
Nodes & Compact & Unique Circuits & Classes \\ \toprule
\multirow{2}{*}{2}& 0 & 2 & 1 \\
& 1 & 2 & 1 \\
\midrule
\multirow{3}{*}{3}& 0 & 44 & 10 \\
& 1 & 26 & 6 \\
& 2 & 10 & 3 \\
\midrule
\multirow{4}{*}{4}& 0 & 6824 & 122 \\
& 1 & 3108 & 74 \\
& 2 & 892 & 28 \\
& 3 & 158 & 8 \\
\midrule
\multirow{5}{*}{5}& 0 & 5871560 & 2869 \\
& 1 & 1669398 & 1366 \\
& 2 & 363929 & 509 \\
& 3 & 66683 & 125 \\
& 4 & 5379 & 24 \\
\bottomrule
\end{tabular}
\caption{Number of unique circuits and Hamiltonian classes for circuits with the specified numbers of nodes and compact variables. As node count increases, compact variables appear in fewer unique circuits but in a larger fraction of Hamiltonian classes.}
    \label{tab:stats_periodic}
\end{table}

\newpage
\begin{table}[ht]
\centering
\setlength{\tabcolsep}{13pt}
\begin{tabular}{cccc}
\toprule
Nodes & Loops & Unique Circuits & Classes \\ \toprule
\multirow{2}{*}{2}& 0 & 2 & 1 \\
& 1 & 2 & 1 \\
\midrule
\multirow{5}{*}{3}& 0 & 12 & 4 \\
& 1 & 28 & 8 \\
& 2 & 24 & 7 \\
& 3 & 12 & 3 \\
& 4 & 4 & 1 \\
\midrule
\multirow{10}{*}{4}& 0 & 340 & 18 \\
& 1 & 1163 & 62 \\
& 2 & 2098 & 94 \\
& 3 & 2585 & 87 \\
& 4 & 2292 & 62 \\
& 5 & 1491 & 39 \\
& 6 & 708 & 20 \\
& 7 & 238 & 8 \\
& 8 & 56 & 2 \\
& 9 & 11 & 1 \\
\midrule
\multirow{17}{*}{5}& 0 & 12826 & 84 \\
& 1 & 68702 & 501 \\
& 2 & 227894 & 1273 \\
& 3 & 477011 & 1811 \\
& 4 & 897840 & 1972 \\
& 5 & 1230312 & 1831 \\
& 6 & 1431401 & 1546 \\
& 7 & 1365488 & 1191 \\
& 8 & 1062982 & 844 \\
& 9 & 666862 & 532 \\
& 10 & 338594 & 305 \\
& 11 & 138860 & 150 \\
& 12 & 44599 & 68 \\
& 13 & 11144 & 25 \\
& 14 & 2104 & 7 \\
& 15 & 296 & 2 \\
& 16 & 34 & 1 \\
\bottomrule
\end{tabular}
\caption{Number of unique circuits and Hamiltonian classes by number of flux loops. Circuits with different numbers of flux loops can belong to the same Hamiltonian class.}
    \label{tab:granular_overview2}
\end{table}

Two points follow from \cref{tab:granular_overview2}. First, circuits with different numbers of flux loops can belong to the same Hamiltonian class. Second, in three-node circuits, inductors appear to correspond to extended or harmonic modes, up to the total number of modes present, while a Josephson junction without an accompanying inductor is needed to produce a compact mode. We further explore the impact of the inductive portion of the circuit in the next section.

\section{Inductive Skeletons with Delocalized Modes}\label{app:skeleton}

In this appendix, we define the delocalized coordinate used in \cref{sec:delocalized} and derive necessary and sufficient conditions for its existence.
The criteria depend only on the \emph{inductive skeleton}, the subcircuit of junctions and inductors, so can be applied before choosing any capacitances.

\subsection{Conditions for Delocalized Modes}

Throughout this section, an admissible coordinate transformation satisfies the mode-separation and compact-lattice requirements of \cref{app:var_trans}. The results below apply only to such transformations.

We call an extended coordinate $\tilde\Phi_i$ \emph{localized} if it can be chosen as the phase across a single Josephson junction. That junction phase is $\pm\tilde\Phi_i$ and contains no other coordinate. 
An extended coordinate is \emph{delocalized} if no admissible transformation allows such a choice. Our goal is to determine whether a given inductive skeleton requires delocalized coordinates and, if so, how many.

We first count the compact and extended coordinates. Let $\omega_J$ and $\omega_L$ denote the junction and inductor incidence matrices for an $N$-node circuit, and define
\begin{subequations}
\begin{align}
    &r_J = \textrm{rank}\,\omega_J, \quad r_L = \textrm{rank}\,\omega_L \, ,
    \label{eq:skel_ranks}\\
    &r_{JL} = \textrm{rank}\,[\,\omega_J\ \ \omega_L\,] \leq r_J + r_L\, .
\end{align}
\end{subequations}
Here the rank is the number of linearly independent columns. We also define two nullspaces,
\begin{align}
    J= \textrm{null}(\omega_J^T) \, , \qquad L= \textrm{null}(\omega_L^T) \, .
    \label{eq:skel_nullspaces}
\end{align}
 Thus $J$ contains the node-flux directions that change no junction phase, while $L$ contains those that produce no inductive energy.

A compact coordinate lies in $L$, because it has no quadratic inductive energy. The intersection $J\cap L$ contains directions seen by neither junctions nor inductors. We therefore define the compact subspace $\mathcal{C}\equiv L\cap(J\cap L)^{\perp}$, where $A^{\perp}$ denotes the directions orthogonal to every vector in $A$. 
Thus $\mathcal{C}$ is the part of $L$ orthogonal to these shared null directions, as illustrated in \cref{fig:mode_diagram}. Since $\dim J=N-r_J$, $\dim L=N-r_L$, and $\dim(J\cap L)=N-r_{JL}$, the number of compact coordinates is
\begin{align}
    N_{\rm c}=\dim\mathcal{C}=r_{JL}-r_L \, .
\end{align}
The total number of nonlinear coordinates equals the number of independent junction branches. Thus $N_{\rm c}+N_{\rm e}=r_J$ where $N_{\rm e}$ is the number of extended nonlinear coordinates, giving
\begin{align}
    N_{\rm e}=r_J+r_L-r_{JL} \, .
\end{align}
This count assumes the intrinsic junction capacitance used throughout this paper. The inductive skeleton fixes the compact/extended split, and added capacitances may change the remaining $N-r_J$ coordinates among harmonic, free, frozen, and cyclic types.

We call a junction \emph{inductively shunted} if an inductor-only path connects its two nodes. Equivalently, its incidence column lies in $\textrm{col}\,\omega_L$, the column space of $\omega_L$. Let $\mathcal{S}$ be the set of inductively shunted junctions, and $\omega_{J,\mathcal{S}}$ be the matrix formed from their incidence columns.

To express the localization condition algebraically, write the node fluxes as $\Phi=Z\tilde\Phi$, where the columns $z_j$ of $Z$ define the new coordinates. The phase across junction $k$ is
\begin{align}
    \omega_{J,k}^T\Phi
    =\sum_j\left(\omega_{J,k}^Tz_j\right)\tilde\Phi_j \, .
    \label{eq:skel_jphase}
\end{align}
The coordinate $\tilde\Phi_i$ is localized to junction $k$ if this sum contains only $\pm\tilde\Phi_i$. 
Equivalently,
\begin{align}\label{eq:localized}
    \omega_{J,k}^TZ=\pm e_i^T
    \quad\Longleftrightarrow\quad
    \omega_{J,k}=\pm (Z^{-1})_{i}^{T} \, ,
\end{align}
where $e_i$ is the $i$th standard basis vector and $(Z^{-1})_{i}$ is the $i$th row of $Z^{-1}$.
The columns of $Z$ corresponding to compact, free, and cyclic coordinates span $L$. By \cref{eq:localized}, $\omega_{J,k}$ is orthogonal to all of these columns. Therefore 
$\omega_{J,k} \in L^\perp = \textrm{col}(\omega_L)$, and junction $k$ must be inductively shunted.
Distinct localized extended coordinates correspond to linearly independent rows of $Z^{-1}$, so at most $\textrm{rank}\,\omega_{J,\mathcal{S}}$ can be localized simultaneously.

This upper bound is always attainable. Let $d=\textrm{rank}(\omega_{J,\mathcal S})$ and select $d$ independent shunted-junction incidence columns. Each selected column is orthogonal to both $J$ and $L$. The $N_{\rm e}$ extended rows of $Z^{-1}$, after transposition, form a basis for the $N_{\rm e}$-dimensional subspace orthogonal to both $J$ and $L$. We may therefore choose the selected incidence columns as the first $d$ transposed extended rows and complete them to a basis. This changes the basis only within the extended sector and therefore neither mixes mode types nor changes the compact lattice. By \cref{eq:localized}, the first $d$ extended coordinates are then phases across the selected junctions. The bound is therefore attained, and all extended coordinates can be localized if and only if $\textrm{rank}\,\omega_{J,\mathcal{S}}=N_{\rm e}$. The number that must remain delocalized is
\begin{align}
    \delta=N_{\rm e}-\textrm{rank}\,\omega_{J,\mathcal{S}} \, .
    \label{eq:skel_delta}
\end{align}
The index $\delta$ counts only extended coordinates. Compact coordinates are handled by the lattice construction of \cref{app:var_trans}.

The criterion also has a simple graphical interpretation. An extended coordinate arises from a loop containing both junctions and inductors. If such a mixed loop contains only one junction, the remaining inductors form an inductor-only path across that junction, so the coordinate can be localized to that junction. 
A delocalized coordinate therefore requires a mixed loop with at least two junctions. If no junction is inductively shunted, such a loop is also sufficient.

A nonzero $\delta$ means that the circuit must contain at least one delocalized nonlinear coordinate. It does not imply delocalized eigenstates or protection from noise; these require parameter-dependent spectral and noise analysis. Nevertheless, protected circuits motivate using $\delta>0$ as a topology-level design filter \cite{Gyenis2021,kalashnikov_bifluxon_2020}. The inductive skeleton fixes the nonlinear mode structure, but not the protected regime; \cref{fig:copycat_qubits} shows that circuits in the same Hamiltonian class need not share the same protected limit.
Circuits with localized modes can be protected. For example the Josephson rhombus underlying the protected chains of Ref.~\cite{Bell2014} consists of nonlinearly coupled compact modes. Similarly, our classification identifies the Difluxmon qubit \cite{garcia-azorin_2024} as relying on coupled localized modes to balance protection and control.
Thus $\delta$ screens for collective nonlinear structure, not qubit protection.

\subsection{Catalog and Construction of Skeletons}

We evaluate \cref{eq:skel_delta} for every enumerated inductive skeleton and show the unique irreducible-skeletons in \cref{fig:nontrivial_classes}. We call a delocalized skeleton irreducible if $\delta = N_e$ and it meets the following conditions. First, its circuit graph is biconnected: removing any single node does not disconnect the remaining graph. Next, we exclude skeletons with an inductor directly in parallel to a junction or an internal node connected only by inductors. We also exclude all-junction or all-inductor loop additions that leave $\delta$ unchanged. A non-biconnected skeleton separates into smaller blocks, and both terms in \cref{eq:skel_delta} split across these blocks, so their contributions to $\delta$ add. Similarly, branches attached through a single connection can add localized compact, harmonic, or non-dynamical coordinates, but do not create a new delocalized mode. Capacitors between different blocks may still couple their dynamics. The figure is therefore a catalog of irreducible-skeletons, not of every skeleton with $\delta\geq1$.

One irreducible-skeleton appears at three nodes: the skeleton of class $3-16$ shown in \cref{fig:ind_cir_three}(d). Four appear at four nodes and seventeen at five nodes, giving the twenty-two examples in \cref{fig:nontrivial_classes}. The same filters yield $149$ six-node irreducible-skeletons, so we stop the catalog at five nodes and provide the general criterion instead.

Larger skeletons can be built from the irreducible-skeletons in three ways. First, a branch can be extended in series by inserting a node and an additional element. For example, the KITE element \cite{smith2020TwoCooperPairTunneling} can be constructed from the \texttt{J}--\texttt{L}--\texttt{J}--\texttt{L} loop of \cref{fig:nontrivial_classes}. This loop is also the inductive skeleton of the \zp\ circuit. The KITE is obtained by replacing each inductor branch with two inductors in series. Second, a branch or chain can be added between existing nodes to create a new loop. Repeating these two operations generates larger skeletons. Each candidate must still be checked using \cref{eq:skel_delta}, because a new inductive path can shunt a junction and thereby localize a coordinate. Third, two skeletons can be joined at a single node. Their values of $\delta$ then add block by block. Such joined structures are composite rather than irreducible and are therefore omitted from the catalog, although capacitances may still couple the dynamics of the different blocks.

All skeletons are nevertheless present in the full enumeration database. Through an inverse lookup as demonstrated in \cref{sec:lookup}, all derived capacitive configurations could be identified and used as a starting point for an explicit protected qubit search.

\section{Modeling Implementation Details}\label{app:details}
In modeling previously unstudied circuits, we are forced to make several key assumptions. Many are taken for granted in transmon-like systems, but they can have a much larger effect in multimode circuits. We list the main assumptions below; the list is not exhaustive.

\textbf{Selection of computational states}. The computational states are taken to be the ground state and first excited states: $\ket{0}_L = \ket{0}$ and $\ket{1}_L = \ket{1}$. As outlined in \cref{app:gate_time}, the dominant state for leakage errors is assumed to be the second excited state $\ket{2}$.

\textbf{$1/f$ Noise Amplitudes.} We use the \verb|SQcircuit| default values of $10^{-4}$, $10^{-6}$, and $10^{-7}$ for charge, flux, and critical current noise respectively. These values align with the lower bounds laid out in \cite{houck_life_2009}.

\textbf{Hilbert-space cutoff}. \verb|SQcircuit| represents compact modes in a charge basis and extended modes in a Fock basis. To choose the cutoff for each mode, we randomly initialize 200 circuits within the parameter search space. For each circuit, we increase the cutoff on each mode by a fixed increment until the fractional change in the first five eigenvalues falls below $10^{-6}$. To balance speed and accuracy, we then choose, for each circuit, the 90th-percentile cutoff over the sampled points. We finally re-evaluate each optimum using a cutoff chosen for that specific parameter set. Only 2 of the 84 circuits showed a meaningful change in the predicted  $N_{\rm gate}$ after re-evaluation.

\textbf{Optimizing offsets}. Because circuit performance can depend strongly on external inputs, we treat flux and charge offsets as discrete variables. They may take the values $0$, $0.25$, $0.5$, and $0.75$ [$\phi_0$ for flux], shifted slightly by $10^{-4}$ for charge and $10^{-6}$ for flux. These shifts avoid numerical instabilities at exactly $0$ or $0.5$ and represent a worst-case DC offset error equal to the $1/f$ noise amplitudes used in the $T_\phi$ calculation.

\textbf{External-flux gauge}. For each enumerated circuit, we construct an independent cycle basis and associate an \texttt{SQcircuit} loop object with each cycle. Given these loops and the full capacitance network, \texttt{SQcircuit} determines the parameter-dependent irrotational distribution of external flux over the circuit branches~\cite{YouKoch2019,BryonExtFlux2023,Rajabzadeh2023analysisofarbitrary}. This satisfies the loop-flux constraints while eliminating the spurious coupling proportional to $\dot{\Phi}_{\rm ext}$, and is recalculated for the circuit parameters used in each numerical analysis.

\textbf{Coupling}. We assume that each mode in a multimode circuit can be coupled independently through its offset charge or flux. We make this assumption so that circuit performance can be optimized over those offsets. We also ignore coupling capacitances associated with charge offsets when constructing the Hamiltonian.

\textbf{Parameter values}. We limit the underlying circuit parameters to slightly beyond experimentally realized values. We consider Josephson junctions to have a fixed small capacitance. See \cref{tab:params_overview} for exact values.

\textbf{Ground placement}. The choice of ground can significantly affect circuit performance. We model each node as weakly capacitively coupled to ground. An alternative would be to ground individual nodes explicitly and then use the component-graph isomorphism method of \cref{sec:equiv_reduction} to identify unique ground placements.

\textbf{Computation Time}. To optimize all three-node circuits, we used about 30,000 core-hours on a server with dual AMD EPYC 7713 processors and 2 TB of RAM. Most of this time was spent on circuits with two extended modes, which require much larger Hilbert-space cutoffs. Optimizing a single three-node circuit should be feasible on a typical workstation, although optimizing the full set requires greater computational resources. The full optimization pipeline is included in our repository \cite{sircuitenum_repo}.

\section{Converting Between Physical and Frequency Units for Circuit Parameters}\label{app:params}
Consider dimensionful ($\Phi$, $Q$) and dimensionless ($\varphi$, $q$) flux and charge operators
\begin{align}
    q = \frac{Q}{2e}, \quad {\rm and} \quad 
    \varphi = \frac{2\pi}{\Phi_0}\Phi.
\end{align}
Here $e$ is the charge of a single electron, and $\Phi_0$ is the magnetic flux quantum. $E_C$ \& $E_L$ for a single element are related to physical capacitance $C$ and inductance $L$ by
\begin{align}
 h E_C = \frac{e^2}{2C}, \quad \& \quad
 h E_L = \frac{(\Phi_0/2\pi)^2}{L}\,.
\end{align}
These definitions ensure that the energy stored in a capacitor with capacitance $C$ and an inductor with inductance $L$
\begin{align}
 4 E_C q^2 = \frac{1}{2C}Q^2, \quad \& \quad 
 \frac{1}{2} E_L \varphi^2 = \frac{1}{2 L} \Phi^2   
\end{align}
is consistent between dimensionless and dimensionful units.

\section{Gate Time Estimates}\label{app:gate_time}

\begin{figure}[ht]
\centering 
\includegraphics[width=0.99\columnwidth]{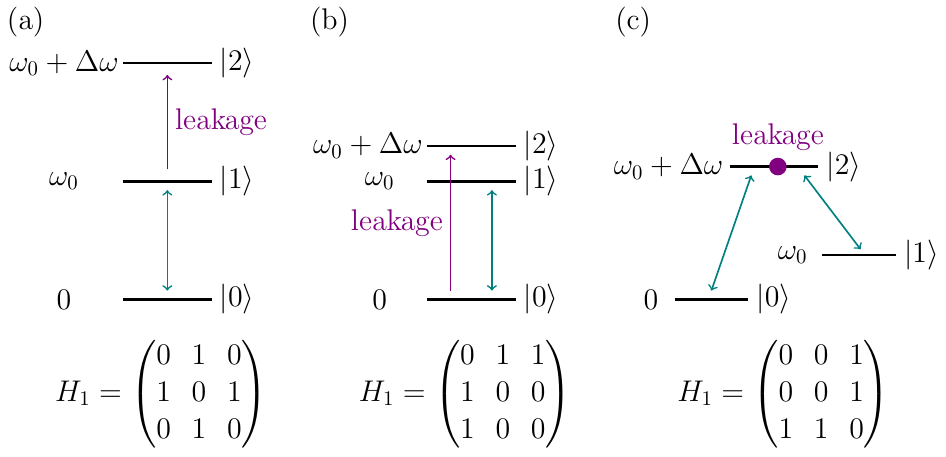}
\caption{Transitions and leakage channels included in the gate-time estimate. (a) A weakly anharmonic qubit driven directly on the $\ket{0} \rightarrow \ket{1}$ transition, with leakage through $\ket{1} \rightarrow \ket{2}$. (b) A qubit with a nearly degenerate second excited state, again driven directly on $\ket{0} \rightarrow \ket{1}$, with leakage through $\ket{0} \rightarrow \ket{2}$. (c) A Raman-type transition $\ket{0} \rightarrow \ket{2} \rightarrow \ket{1}$, where population in $\ket{2}$ is treated as leakage.} \label{fig:gate_times}
\end{figure}

We estimate the gate time by modeling an arbitrary superconducting qubit as a three-level system. We consider either a direct transition, $\ket{0} \leftrightarrow \ket{1}$, or an indirect transition through the higher-lying state $\ket{2}$, that is, a $\Lambda$-type configuration; see \cref{fig:gate_times}. In larger circuits, either transition may require difficult multi-node couplings. The main purpose of this appendix is to treat leakage in a simple and consistent way for both direct and Raman-type transitions. We therefore compare the time required to transfer population from $\ket{0}$ to $\ket{1}$ using both schemes, subject to leakage and drive constraints, and take the faster of the two as the gate-time estimate.

\subsection{Direct Transition}
We model the qubit as a three-level system with Hamiltonian
\begin{align}
    H = H_0 + f(t)H_1 ,
\end{align}
where the free Hamiltonian is
\begin{align}
H_0 =
\begin{pmatrix}
    0 & 0 & 0 \\
    0 & \omega_0 & 0 \\
    0 & 0 & \omega_0 + \Delta\omega
\end{pmatrix} \, .
\end{align}
Here $|\Delta\omega|\geq 0$ and $\omega_0 \geq 0$, and we set the ground-state energy to zero. In \cref{fig:gate_times}, case (a) uses the drive Hamiltonian
\begin{align}
H_1 =
\begin{pmatrix}
    0 & 1 & 0 \\
    1 & 0 & 1 \\
    0 & 1 & 0
\end{pmatrix} \, .
\end{align}
The upper-left $2\times 2$ block drives the qubit $\sigma_X$ transition, while the remaining term provides a leakage channel. The relevant leakage path depends on which transition, $\ket{0}\leftrightarrow\ket{2}$ or $\ket{1}\leftrightarrow\ket{2}$, lies closer in frequency to the qubit transition $\ket{0}\leftrightarrow\ket{1}$. 

We take the drive to be a single-tone field at frequency $\omega_0$ with a Gaussian pulse envelope $f(t) = V_0 e^{-t^2/2\tau^2}\cos(\omega_0 t)$. Here $V_0$ is the peak drive amplitude. In general, the matrix elements of $H_1$, taken here to be unity, depend on the coupling scheme of the circuit.

We expect population to transfer between $\ket{0}$ and $\ket{1}$ when the pulse area is $\pi$. This corresponds to a pulse half-width of $\tau = \sqrt{\pi/2}/V_0$. By setting a maximum value for $V_0$, we obtain a minimum gate time. Faster gates, however, also increase leakage. A shorter pulse has a broader frequency spectrum, which drives the leakage transition more strongly. For weakly anharmonic systems such as the transmon, this is the main source of leakage. The DRAG protocol can suppress this effect \cite{Motzoi_DRAG_2009,Gambetta_DRAG_2009}.

To estimate leakage, we study the effective Hamiltonian:
\begin{align}
    H_\textrm{2-tone} = \begin{pmatrix}
    0 & 1 & 0 \\
    1 & 0 & \lambda \\
    0 & \lambda & 0
\end{pmatrix}\quad
\textrm{or}\quad
\begin{pmatrix}
    0 & 1 & \lambda \\
    1 & 0 & 0 \\
    \lambda & 0 & 0
\end{pmatrix}\, ,
\end{align}
with $\lambda = |F(\omega_0+\Delta\omega)/F(\omega_0)|$ or $\lambda = |F(\Delta\omega)/F(\omega_0)|$, the amplitude ratio of the Fourier transform of the drive at the leakage transition compared with the qubit transition. In chapter 9.5 of  Ref.~\cite{stancil2022principles}, it is shown that for our choice of  $f(t)$, we have
\begin{align}
    F(\omega) = \sqrt{\frac{\pi}{2}}\tau V_0(e^{-(\omega-\omega_0)^2 \tau^2/2} +e^{-(\omega+\omega_0)^2 \tau^2/2}) \, . 
\end{align}
The amplitude ratio is thus $\lambda = e^{-\delta^2 \tau^2/2}$, where $\delta$ is the detuning between the leakage transition and the qubit transition (either $\Delta\omega-\omega_0$ or $\Delta\omega$ depending on the level structure).

Solving the time-dependent Schr\"odinger equation with $\ket{\psi(t=0)} = \ket{0}$, the maximum probability to be in the $\ket{1}$ state (in both cases) is
\begin{align}
    P_{\max,\ket{1}} = \frac{1}{1+\lambda^2} = \frac{1}{1+e^{-\delta^2\tau^2}} \, .
\end{align}
To estimate the gate time, we substitute $\tau = \sqrt{\pi/2}/V_0$ corresponding to a $\pi$ pulse and solve for the $V_0$ required to achieve a population transfer of 99.9\%
\begin{align}
    \frac{1}{1+e^{-\delta^2(\sqrt{\pi/2}/V_0)^2}} = 0.999 \rightarrow V_0 \approx \frac{\delta}{2} \, .
\end{align}

Finally, we impose two drive limits. First, we assume an absolute lower bound of about 1 ns for the $\pi$-pulse width, which gives a maximum drive amplitude $V_{\rm max} = 200$ MHz. Second, we require $10 V_0 \leq \omega_0$, so that the direct drive remains within the regime described above. This gives
\begin{align}\label{eq:gtime}
\tau_{g,\textrm{direct}} =
    \begin{cases} 
      \sqrt{2\pi}/\delta, & \delta/2 \leq V_{\rm lim}\\
      \sqrt{\pi/2}/V_{\rm max} ,& \delta/2 > V_{\rm lim}
   \end{cases} \, ,
\end{align}
with $\delta$ again being the detuning between $\omega_0$ and the leakage transition ($\ket{0}\leftrightarrow\ket{2}$ or $\ket{1}\leftrightarrow\ket{2}$) and $V_{\rm lim} = \min(V_{\rm max}, \omega_0/10)$. Note that reported gate times are set to $5 \tau$, which corresponds to a pulse area of $\approx 99 \%$ of the Gaussian envelope. We see three distinct limiting factors emerge from \cref{eq:gtime}: (1) $\tau$ increases when $\delta$ is small, (2) $\tau$ increases when $\omega_0$ is small, and (3) the absolute speed limit (i.e., $\tau_{\rm min}$ is set by $V_{\rm max}$). In the case of transmon-like systems, factor (1) reproduces the conventional wisdom that gate time is inversely proportional to anharmonicity.

\subsection{Raman-Type Gate}

For transitions via $\ket{2}$, we take $H_0$ to be identical to the direct transition case, but we now consider
\begin{align}
H_1 = \begin{pmatrix}
    0 & 0 & 1 \\
    0 & 0 & 1 \\
    1 & 1 & 0
\end{pmatrix}
\end{align}
and drive the $\ket{0}\leftrightarrow\ket{2}$ and $\ket{1}\leftrightarrow\ket{2}$ transitions with
\begin{align}
    f(t) = e^{-t^2/2\tau^2}[V_1\cos(\omega_0 t - \Delta) + V_2\cos(\Delta\omega t - \Delta)] \, .
\end{align}
When the state begins in $\ket{\psi(t=0)} = \ket{0}$, it experiences Rabi oscillations between $\ket{0}$ and $\ket{1}$ with an effective Rabi frequency of $\Omega_R = V_1V_2/2\Delta$ ~\cite{Gyenis2021}. The intermediate state population experiences oscillations with a maximum value of 
\begin{align}
    P_{\max,\ket{2}} = \frac{V_1^2}{\Delta^2 + V_1^2 + V_2^2} \, .
\end{align}
Taking this maximum intermediate state population as a proxy for leakage, we find the $\tau$ required to limit leakage to 0.1\% (equivalent to the 99.9\% population transfer considered in the direct case). Taking $V_1 = V_2 = V$ and $\Delta = 30 V$ yields an effective Rabi frequency of $\Omega_R = V/60$, with corresponding $\tau = 60\sqrt{\pi/2}/V$ for a $\pi$ pulse. We assume here that there exists a spectrally isolated higher level state, so our only constraint is that $V \leq V_{\rm max}$. This gives $\tau$ for a Raman-style transition of
\begin{align}
    \tau_{g,\textrm{raman}} = 60\sqrt{\pi/2}/V_{\rm max}
\end{align}
Taking $V_{\rm max} = 200 \textrm{ MHz}$ as in the direct transition, we get $\tau_{g,\textrm{raman}} \approx 60$ ns. Again, the gate time is taken to be $5\tau$ to capture most of the pulse area. While $V_{\rm max}$ in the direct transition established a minimum gate time, the same constraint here establishes a maximum gate time of approximately $300$ ns.

The Raman-type gate will be preferred over the direct transition only in the cases of nearly degenerate ground states and very small anharmonicity. Due to the slow nature of these estimates, we do not consider leakage due to spectral broadening for the Raman-style transition.

\section{Comparing Model with Published Results}\label{app:sweeps}

To validate our model of qubit performance (e.g., \cref{eq:ngate}), we compare our predicted optimal operating points for the transmon (\texttt{JC}) and fluxonium (\texttt{JCL}) circuits with published experimental values. Including charge offset $n_g$ and external flux $\varphi_{\rm ext}$, the Hamiltonians for the two circuits are
\begin{align}
    \label{eq:tmon_flux_ham}
    H_{\texttt{JC}} &= 4E_C (n-n_g)^2 - E_J \cos(\theta) \\
    H_{\texttt{JCL}} &= 4E_C q^2 +\frac{1}{2}E_L \varphi^2 - E_J \cos(\varphi - \varphi_{\rm ext}) \,.
\end{align}

\begin{figure}[ht]
\centering 
\includegraphics[width=0.99\columnwidth]{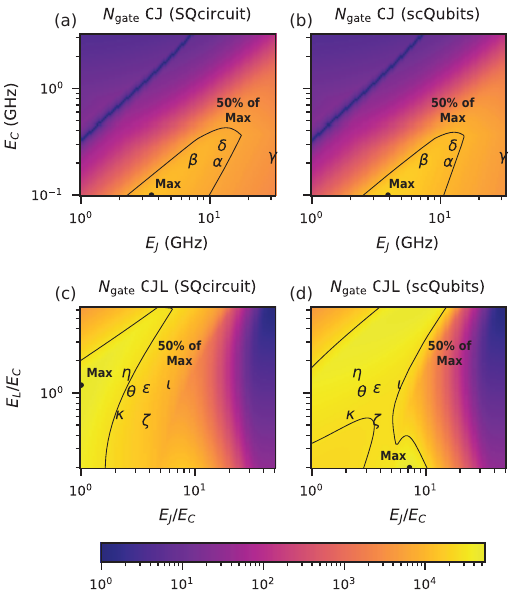}
\caption{Qualitative estimate of $N_{\rm gate}$ calculated for the \texttt{JC} (a-b) and \texttt{JCL} (c-d) circuits. The left column (a,c) uses the \texttt{SQcircuit} noise model, which uses a second-order approximation for pure dephasing, while the right column (b,d) uses the \texttt{scQubits} noise model, which uses a first-order approximation for pure dephasing. For the \texttt{JC} circuit, $n_g = 0.25$ is chosen to represent a worst-case when a DC offset is not included. For the \texttt{JCL} circuit, the standard $\varphi_{\rm ext} = \pi$ is used, and $E_C$ is fixed at 500 MHz. The black dot marks the maximum point, and the black contour shows the region that is at least 50\% of the maximum value. The maximum value of $N_{\rm gate}$ is approximately 12,500 for the transmon (\texttt{JC}) and 54,500 for fluxonium (\texttt{JCL}). The streak with low $N_{\rm gate}$ for the \texttt{JC} circuit is due to a zero crossing of the anharmonicity. The bottom right represents the Transmon regime ($E_J/E_C \gg 1$), while the top left represents the Cooper Pair Box, or charge qubit, regime (small $E_J/E_C$). To gauge the discrepancies between our model and experimental data, we extracted the circuit parameters from various published qubit designs. Each letter corresponds to parameters from a published paper: transmon $\alpha$--\cite{krinner_realizing_2022}, $\beta$--\cite{manenti_full_2021}, $\gamma$--\cite{arute_quantum_2019}, and $\delta$--\cite{berke_transmon_2022}; fluxonium $\epsilon$--\cite{nguyen_high-coherence_2019}, $\zeta$--\cite{zhang_universal_2021}, $\eta$--\cite{bao_fluxonium_2022}, $\theta$--\cite{somoroff_fluxonium_2024}, $\iota$--\cite{somoroff_millisecond_2023}, and $\kappa$--\cite{moskalenko_high_2022}. Qubits from the literature fall near or within the best performing regions for both the transmon and fluxonium circuits.}
\label{fig:ngate_sweep}
\end{figure}

We visualize $N_{\rm gate}$ as a 2-D parameter sweep in \cref{fig:ngate_sweep}. In the case of fluxonium, $E_C$ is fixed at 500 MHz, with $E_L$ and $E_J$ expressed in units of $E_C$. For the transmon, $n_g = 0.25$ is chosen to represent a worst-case offset when a DC offset is not included. For fluxonium, the standard $\varphi_{\rm ext} = \pi$ is used. To provide a fair comparison with the published experimental values, we restrict our parameter ranges slightly relative to the ranges in \cref{tab:params_overview} and consider only capacitors and inductors of $E_C$ and $E_L \geq 0.1$ GHz. We use two popular software packages which have different noise models, namely \texttt{SQcircuit} \cite{Rajabzadeh2023analysisofarbitrary,rajabzadeh_2024} and \texttt{scQubits} \cite{Groszkowski2021_scqubitspython, chitta2022_scqubits}.

In \cref{fig:ngate_sweep} (a-b) the published transmon parameters ($E_J$ and $E_C$) from the literature fall close to the 50\% contour of the numerically optimal point. The agreement is remarkable, perhaps reflecting that noise in the transmon is well understood. In the case of fluxonium, shown in \cref{fig:ngate_sweep} (c-d), the situation is more complicated.
When using the \texttt{SQcircuit} package
fewer of the published experiments fall within 50\% of our numerically optimized point. When using the \texttt{scQubits} package, the agreement improves dramatically. We believe the difference between the packages' models arises from a difference in pure dephasing rates. \texttt{SQcircuit} uses a second-order approximation, while \texttt{scQubits} uses a first-order approximation. As a result, the \texttt{SQcircuit} noise model penalizes larger $E_J$ values more due to second-order critical current noise (see \cref{fig:trans_flux_decay} for a breakdown of the individual noise channels). Nevertheless, all points except $\iota$ fall above 25\% of the maximum value for the \texttt{SQcircuit} noise model. The agreement is still impressive considering neither the noise nor gate speed models have been fine-tuned for fluxonium.

As an additional validation of the gate-time estimate in \cref{app:gate_time}, we compare our predicted gate times with the experimentally reported gate times of the reference transmon and fluxonium devices. The data are listed in \cref{tab:paper_data_tmon} and \cref{tab:paper_data_flux}. The model reproduces the ordering of the reported gate times for all devices that report one, although the variation is much smaller for the transmons. It does not predict the times exactly, but it appears to serve its intended role as a high-throughput speed limit for the optimization procedure. Somewhat surprisingly, the transmon estimates disagree more with experiment than the fluxonium estimates. One likely reason is that our target single-qubit gate coherent fidelity of 
99.9\% is lower than the 
99.99\% target relevant for the transmon devices. The context also differs: the transmon devices were integrated into a larger architecture (perhaps with e.g. DRAG pulses), whereas the fluxonium devices were more isolated research devices.

\newpage
\begin{widetext}

\begin{table}[ht]
\setlength{\tabcolsep}{3.5pt}
\footnotesize
\begin{tabular}{@{}lccccccccccc@{}}
\toprule
Group  or Device &  $E_J$ (GHz) &  $E_C$ (GHz) & $f_{01}, \eta$ (GHz) & $T_1$ ($\mu$s)& $T_2$ ($\mu$s)& $T_2^{\rm echo}$ ($\mu$s)  & $T_{\rm gate}$ (ns) & $T_{\rm gate, est}$ (ns) & $N_{\rm gate}$ \\ \toprule

$\alpha$--\cite{krinner_realizing_2022} ETH Zurich (2022) & 10.5 & 0.18  &  3.7, -0.18     &  60 &  78 & 89  & 40   & 11 & 1950  \\\midrule

$\beta$--\cite{manenti_full_2021} Rigetti (2021)   & 6.8  & 0.18  &  3.4, -0.21  &  51 &  10 & 81   & 20  & 9 &  500 \\\midrule

$\gamma$--\cite{arute_quantum_2019} Google Sycamore (2019)  & 28.3  & 0.21  &  6.7, -0.21     &  16 &  -- & --    & 20 & 9 & 1280 \\\midrule

$\delta$--\cite{berke_transmon_2022} IBM Brooklyn (2022)  & 11.5  & 0.25  &  --     &  -- &  -- & --    & --    & -- & --   \\\midrule

\end{tabular}
    \caption{Transmon: estimates of device parameters shown in \cref{fig:ngate_sweep}, extracted from published results. If incomplete parameters are given, we infer missing values by approximating $\eta \approx -E_C$ and $f_{01} \approx \sqrt{8 E_C E_J} - E_C$. For tunable qubits, if no idle frequency is given, we use the middle of the given range. $N_{\rm gate}$ is calculated by $N_{\rm gate} = T_2/T_{\rm gate}$. If no $T_2$ is provided, we take $1/T_2 = 1/2T_1$ -- which implies (incorrectly) that $T_{\phi}$ is infinite. For $\alpha$, we choose device D6, for $\beta$ we choose device q0, and for $\gamma$ we use the mean device parameters.}
    \label{tab:paper_data_tmon}
\end{table}

\begin{table}[ht]
\setlength{\tabcolsep}{2pt}
\footnotesize
\begin{tabular}{@{}lcccccccccc@{}}
\toprule
Group  or Device &  $E_J$ (GHz) &  $E_C$ (GHz) & $E_L$ (GHz) & $f_{01}, \eta$ (GHz) & $T_1$ ($\mu$s)& $T_2$ ($\mu$s)& $T_2^{\rm echo}$ ($\mu$s)  & $T_{\rm gate}$ (ns) & $T_{\rm gate, est}$ (ns) & $N_{\rm gate}$ \\ \toprule

$\epsilon$--\cite{nguyen_high-coherence_2019} Manucharyan (2019)   & 3.43  & 1.00  &  0.58 & 0.39, 2.84  & 500 &  510 & -- & -- & 25 & --  \\\midrule

$\zeta$--\cite{zhang_universal_2021} Schuster (2021)  &  3.40  & 0.48  &  0.13 & 0.01, 2.96  &  315 &  300 & 100   & 120 & 299 & 2500  \\\midrule

$\eta$--\cite{bao_fluxonium_2022} Alibaba (2022)   & 2.26  & 1.40  &  0.52 & 1.09, 1.93     &  80 &  35 & 30 & 10  & 9 & 3500  \\\midrule

$\theta$--\cite{somoroff_fluxonium_2024} SEEQC (2024)  &  2.50  & 0.89  &  1.14 & 1.25, 1.42  &  77 &  29.5 & 38   & 6   & 8 & 4900  \\\midrule

$\iota$--\cite{somoroff_millisecond_2023} Manucharyan (2023)   & 5.57  & 1.08  &  0.64 & 0.16, 5.57     &  1200 &  1160 & 1550  & 80  & 62  & 18,500  \\\midrule

$\kappa$--\cite{moskalenko_high_2022} Besedin (2022)  &  1.99  & 0.56  &  0.86 & 0.75, 1.05  &  172 &  113 & 146  & 13  & 13 & 8500  \\\midrule

\end{tabular}
    \caption{Fluxonium: device parameters shown in \cref{fig:ngate_sweep}, extracted from published results. $N_{\rm gate}$ is calculated by $N_{\rm gate} = T_2/T_{\rm gate}$. $T_{\rm gate, est}$ represents our estimated gate time, calculated using the model described in \cref{app:gate_time}. For $\epsilon$, we choose device I, for $\eta$ we choose device A, for $\theta$ we choose device 3, and for $\kappa$ we choose device S.}
\label{tab:paper_data_flux}
\end{table}

\begin{figure*}[ht!]
\centering 
\includegraphics[width=0.99\textwidth]{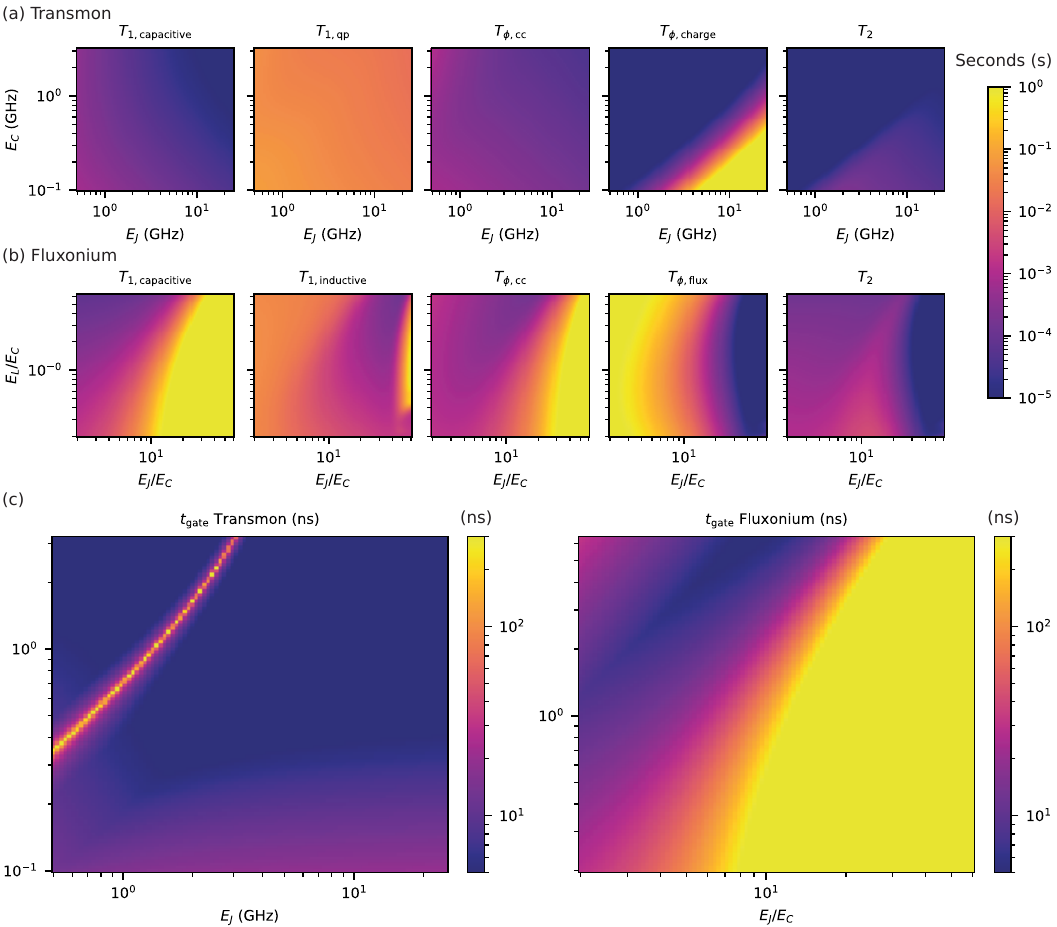}
\caption{Underlying coherence times and gate times estimates for the number of gates $N_{\rm gate}$ estimates in~\cref{fig:ngate_sweep}. Noise estimates from \texttt{SQcircuit} are shown for (a) transmon and (b) fluxonium. Channels not shown did not contribute meaningfully for the respective systems. Gate time estimates based on the analysis given in \cref{app:gate_time} are shown in (c). The gate time estimates for the transmon seem consistent with gate times from the literature. It is important to note that removing noises or reducing the noise magnitude will change the conclusion of our numerics.}
\label{fig:trans_flux_decay}
\end{figure*}

\FloatBarrier
\end{widetext}

\section{Optimization Results by Number of Junctions, Loops, and Compact Variables} \label{app:more_optim_results}

In \cref{tab:optim_jj,tab:optim_charge,tab:optim_loops} we split by JJs, number of charge modes, number of flux loops, respectively. Broadly speaking, these splits tell us about the relative importance of charge and flux noise when modeling the number of gates. From \cref{tab:optim_jj}, we infer that fewer JJs are preferred by our model. From \cref{tab:optim_loops}, we infer that 1 or 2 flux loops are preferred over 0, 3, or 4 loops. These preferences may be a reflection of the limited design space of three-node circuits. More flux loops means a higher sensitivity to flux noise, and with only three nodes, it is difficult to add more junctions without creating new flux loops. In \cref{tab:optim_charge}, we additionally see that compact modes are heavily penalized. Again, this may be because of the limited configurations available in three-node circuits.

\begin{table}[ht]
\setlength{\tabcolsep}{7.5pt}
\centering
\begin{tabular}{cccc}
\toprule
$N_{JJ}$ & Circuit & $N_{\rm gate}$ (Mean) & $N_{\rm gate}$ (Std) \\ \toprule
\multirow{3}{*}{1}& \texttt{J}, \texttt{L}, \texttt{CL} & $2.77\cdot10^{5}$ & $1.6\cdot10^{4}$ \\
& \texttt{J}, \texttt{CL}, \texttt{CL} & $2.50\cdot10^{5}$ & $1.1\cdot10^{4}$ \\
& \texttt{CL}, \texttt{CL}, \texttt{CJ} & $2.38\cdot10^{5}$ & $1.1\cdot10^{4}$ \\
\midrule
\multirow{3}{*}{2}& \texttt{J}, \texttt{L}, \texttt{JL} & $2.39\cdot10^{5}$ & $1.8\cdot10^{4}$ \\
& \texttt{L}, \texttt{JL}, \texttt{CJ} & $2.36\cdot10^{5}$ & $1.1\cdot10^{4}$ \\
& \texttt{C}, \texttt{CJL}, \texttt{CJL} & $1.90\cdot10^{5}$ & $5.5\cdot10^{3}$ \\
\midrule
\multirow{3}{*}{3}& \texttt{CJ}, \texttt{CJL}, \texttt{CJL} & $1.86\cdot10^{5}$ & $3.1\cdot10^{3}$ \\
& \texttt{JL}, \texttt{JL}, \texttt{CJ} & $1.35\cdot10^{5}$ & $2.0\cdot10^{4}$ \\
& \texttt{JL}, \texttt{CJ}, \texttt{CJL} & $1.23\cdot10^{5}$ & $3.4\cdot10^{3}$ \\
\bottomrule
\end{tabular}
    \caption{Optimized number of gates for all three-node circuits, split by number of junctions $N_{JJ}$. Circuits with fewer junctions perform better than those with more junctions.} 
    \label{tab:optim_jj}
\end{table}

\begin{table}[ht]
\setlength{\tabcolsep}{6.4pt}
\centering
\begin{tabular}{cccc}
\toprule
$N_{\rm loop}$ & Circuit & $N_{\rm gate}$ (Mean) & $N_{\rm gate}$ (Std) \\ \toprule
\multirow{2}{*}{0}& \texttt{C}, \texttt{J}, \texttt{CL} & $1.24\cdot10^{5}$ & $4.8\cdot10^{3}$ \\
& \texttt{C}, \texttt{J}, \texttt{L} & $1.22\cdot10^{5}$ & $5.6\cdot10^{3}$ \\
\midrule
\multirow{2}{*}{1}& \texttt{J}, \texttt{L}, \texttt{CL} & $2.77\cdot10^{5}$ & $1.6\cdot10^{4}$ \\
& \texttt{J}, \texttt{CL}, \texttt{CL} & $2.50\cdot10^{5}$ & $1.1\cdot10^{4}$ \\
\midrule
\multirow{2}{*}{2}& \texttt{J}, \texttt{L}, \texttt{JL} & $2.39\cdot10^{5}$ & $1.8\cdot10^{4}$ \\
& \texttt{L}, \texttt{JL}, \texttt{CJ} & $2.36\cdot10^{5}$ & $1.1\cdot10^{4}$ \\
\midrule
\multirow{2}{*}{3}& \texttt{CJ}, \texttt{CJL}, \texttt{CJL} & $1.86\cdot10^{5}$ & $3.1\cdot10^{3}$ \\
& \texttt{JL}, \texttt{JL}, \texttt{CJ} & $1.35\cdot10^{5}$ & $2.0\cdot10^{4}$ \\
\midrule
\multirow{2}{*}{4}& \texttt{CJL}, \texttt{CJL}, \texttt{CJL} & $7.04\cdot10^{4}$ & $10.0\cdot10^{2}$ \\
& \texttt{JL}, \texttt{JL}, \texttt{CJL} & $6.56\cdot10^{4}$ & $6.7\cdot10^{2}$ \\
\bottomrule
\end{tabular}
    \caption{Optimized number of gates for all three-node circuits, split by number of flux loops $N_{\rm loop}$. Circuits with one or two loops score the highest.} 
    \label{tab:optim_loops}
\end{table}

\begin{table}[ht]
\setlength{\tabcolsep}{7.0pt}
\centering
\begin{tabular}{cccc}
\toprule
$N_{\rm compact}$ & Circuit & $N_{\rm gate}$ (Mean) & $N_{\rm gate}$ (Std) \\ \toprule
\multirow{3}{*}{0}& \texttt{J}, \texttt{L}, \texttt{CL} & $2.77\cdot10^{5}$ & $1.6\cdot10^{4}$ \\
& \texttt{J}, \texttt{CL}, \texttt{CL} & $2.50\cdot10^{5}$ & $1.1\cdot10^{4}$ \\
& \texttt{J}, \texttt{L}, \texttt{JL} & $2.39\cdot10^{5}$ & $1.8\cdot10^{4}$ \\
\midrule
\multirow{3}{*}{1}& \texttt{C}, \texttt{J}, \texttt{CJL} & $1.88\cdot10^{5}$ & $4.6\cdot10^{3}$ \\
& \texttt{C}, \texttt{CJ}, \texttt{CJL} & $1.87\cdot10^{5}$ & $3.8\cdot10^{3}$ \\
& \texttt{C}, \texttt{JL}, \texttt{CJ} & $1.87\cdot10^{5}$ & $5.1\cdot10^{3}$ \\
\midrule
\multirow{3}{*}{2}& \texttt{J}, \texttt{J}, \texttt{J} & $6.74\cdot10^{4}$ & $1.9\cdot10^{3}$ \\
& \texttt{J}, \texttt{J}, \texttt{CJ} & $6.03\cdot10^{4}$ & $1.7\cdot10^{3}$ \\
& \texttt{CJ}, \texttt{CJ} & $4.53\cdot10^{4}$ & $6.2\cdot10^{2}$ \\
\bottomrule
\end{tabular}
    \caption{Optimized number of gates for all three-node circuits, split by number of compact modes $N_{\rm compact}$. Circuits with fewer compact modes perform better.} 
    \label{tab:optim_charge}
\end{table}

\section{Circuit Quantization Details for Top Performing Circuit}\label{app:quantization_details}

\begin{figure}[h]
\includegraphics[width=0.5\columnwidth]{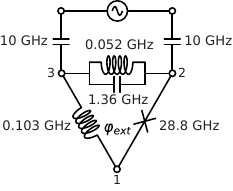}
    \caption{Circuit diagram for the top performing circuit, coupled to an AC voltage source. Each circuit node is labeled 1, 2, or 3, corresponding to node flux variables $\phi_i$.} 
    \label{fig:new_circuit_diagram}
\end{figure}

Here we derive the Hamiltonian for coupling the top-performing three-node circuit from \cref{sec:optimization} to an AC voltage source. The circuit diagram appears in \cref{fig:new_circuit_diagram}, with node labels corresponding to node-flux variables $\phi_i$.
In the node variables, the circuit has a capacitance matrix
\begin{align}
    C = \begin{pmatrix}
        C_J & -C_J & 0 \\
        -C_J & C_J + C_C + C_{23} & -C_{23} \\
        0 & -C_{23} & C_{23} + C_C
    \end{pmatrix}
\end{align}
and inverse inductance matrix
\begin{align}
    L^* = \begin{pmatrix}
        1/L_{13} & 0 & -1/L_{13} \\
        0 & 1/L_{23} & -1/L_{23} \\
        -1/L_{13} & -1/L_{23} & 1/L_{13} + 1/L_{23}
    \end{pmatrix},
\end{align}
where $L_{ij}$ and $C_{ij}$ are the inductance and capacitance between nodes $i$ and $j$. $C_J$ is the intrinsic junction capacitance and $C_C$ is the coupling capacitance to the voltage source. We do not include the effects of stray capacitance to ground, although this is included in the gate simulations of \cref{fig:new_circuit_drive}. Nodes 2 and 3 are connected to fixed-voltage nodes, giving an offset charge of
\begin{equation}
    Q_V = C_VV = \begin{pmatrix}
        0 & 0 \\
        C_c & 0 \\
        0 & C_c \\
    \end{pmatrix}\!\begin{pmatrix}
        v/2 \\
        -v/2
    \end{pmatrix} = \begin{pmatrix}
        0 \\
        C_c v/2 \\
        -C_c v/2
    \end{pmatrix}\!.
\end{equation}
Here we set the two nodes with fixed voltage to have values of $v/2$ and $-v/2$.

We choose new variables $\varphi_1 = \frac{1}{2}(\phi_3-\phi_2)$, $\varphi_2 = \phi_2-\phi_1$, and $\varphi_3 = (\phi_1 + \phi_2+\phi_3)$, which corresponds to a transformation matrix $\phi = Z\varphi$ of
\begin{align}
    Z = \begin{pmatrix}
        -2/3 & -2/3 & 1/3 \\
        -2/3 & \phantom{-}1/3 & 1/3 \\
         \phantom{-}4/3 & \phantom{-}1/3 & 1/3
    \end{pmatrix},
\end{align}
selected by the coordinate generation algorithm. After decoupling non-dynamical modes, the transformed capacitance, inductance, and charge offset matrices are
\begin{subequations}
\begin{align}
    \tilde{C} &= Z^TCZ = \begin{pmatrix}
        4 C_{23} + 2C_c & 0 & 0 \\
        0 & C_J & 0 \\
        0 & 0 & 2 C_c
    \end{pmatrix} \\
    \tilde{L}^* &= Z^TL^*Z = \begin{pmatrix}
        \frac{4(L_{13} + L_{23})}{L_{13}L_{23}} & -\frac{2}{L_{13}} & 0 \\
        -\frac{2}{L_{13}} & \frac{1}{L_{13}} & 0 \\
        0 & 0 & 0
    \end{pmatrix} \\
    \tilde{Q}_V &= Z^T C_V V = \begin{pmatrix}
        -C_c v \\
        0 \\
        0
    \end{pmatrix}.
\end{align}
\end{subequations}
The third variable is a non-dynamical free variable, uncoupled from the rest of the system, and can be ignored. The remaining two-variable $\tilde{C}$ is diagonal and can therefore be inverted trivially, yielding
\begin{align}
    \tilde{C}^{-1} &=\begin{pmatrix}
        \frac{1}{4 C_{23} + 2C_c} & 0 \\
        0 & \frac{1}{C_J}
    \end{pmatrix}.
\end{align}
After performing a Legendre transformation, we obtain the circuit Hamiltonian
\begin{align}
    H = &\frac{1}{2}(Q - \tilde{Q}_V)^T \tilde{C}^{-1}(Q - \tilde{Q}_V) \nonumber \\
        &+\frac{1}{2}\Phi^T\tilde{L}^*\Phi + U_{L}(\Phi, \Phi_{\rm ext}) +  U_{JJ}(\Phi) \nonumber \\
      = & \frac{(Q_1 + C_cV)^2}{8 C_{23} + 4C_c} + \frac{Q_2^2}{2C_J}  + \frac{2(L_{13} + L_{23})\Phi_1^2}{L_{13}L_{23}} \nonumber \\
      &- \frac{2\Phi_1\Phi_2}{L_{13}} + \frac{\Phi_2^2}{2L_{13}} \nonumber \\
      &+ \frac{2\Phi_1-\Phi_2}{L_{13}}\Phi_{\rm ext} - E_J\cos\left (\frac{2\pi}{\Phi_0}\Phi_2\right),
\end{align}
with dimension-full charge and flux variables $Q_i$ and $\Phi_i$, where $\Phi_0$ is the flux quantum. We have chosen to assign all external flux to the inductor between nodes 1 and 3 to satisfy the irrotational constraint \cite{YouKoch2019,BryonExtFlux2023}. We separate the external (drive) Hamiltonian coupled to the voltage source as 
\begin{align}
H_{\rm drive} = \frac{2C_cVQ_1}{8 C_{23} + 4C_c} = \frac{8 e^2 n_gq_1}{{8 C_{23} + 4C_c}}
\end{align}
for a dimensionless charge offset $n_g = C_cV/2e$ and charge operator $q_i = Q_i/2e$.
Here we have dropped the constant offset from the $(C_cV)^2$ term.

By examining the Hamiltonian, we see a few desirable features of this coupling scheme. First, in the ideal case, the coupling circuit drives only the first mode, whose matrix elements are large enough to enable a direct transition between $|0\rangle$ and $|1\rangle$. Second, the coupling capacitance $C_c$ adds mass to the already heavy mode while keeping the second mode light, so it only weakly perturbs the uncoupled Hamiltonian.
Lastly, coupling to the voltage source is strong, so gates can be driven with small values of $C_c V$. For the optimized pulses in \cref{fig:new_circuit_drive}, the amplitude of the control pulse $V$ is on the order of $\mu$V. Taking into account the effects of stray capacitance does mix the coupling between the two modes, but the majority of the effect remains in the first mode. Using the circuit parameters from \cref{fig:new_circuit_drive}, we have $H_{\rm drive} = n_g(g_1q_1 + g_2q_2)$, with $g_1 \approx 2.45$ GHz and $g_2 \approx 0.87$ GHz.

\bibliography{cir_enum}

\end{document}